\documentclass[twocolumn,aps,groupedaddress,nofootinbib]{revtex4}

\usepackage{graphicx}
\usepackage{dcolumn}
\usepackage{bm}
\usepackage{amsmath,amssymb,amsfonts}
\usepackage{color}
\usepackage{braket}
\usepackage{mathtools}

\begin{document}

\title{High-frequency rectification via chiral Bloch electrons}

\author{Hiroki Isobe}
\author{Su-Yang Xu}
\author{Liang Fu}
\affiliation{Department of Physics, Massachusetts Institute of Technology, Cambridge, Massachusetts 02139, USA}

\begin{abstract}
Rectification is a process that converts electromagnetic fields into a direct current. Such a process underlies a wide range of technologies such as wireless communication, wireless charging, energy harvesting, and infrared detection. Existing rectifiers are mostly based on semiconductor diodes, with limited applicability to small voltage or high frequency inputs. Here, we present an alternative approach to current rectification that uses the intrinsic electronic properties of quantum crystals without using semiconductor junctions. We identify a previously unknown mechanism for rectification from skew scattering due to the inherent chirality of itinerant electrons in time-reversal-invariant but inversion-breaking materials. Our calculations reveal large, tunable rectification effects in graphene multilayers and transition metal dichalcogenides.  Our work demonstrates the possibility of realizing high-frequency rectifiers by rational material design and quantum wavefunction engineering.
\end{abstract}

\maketitle

\subsection*{Introduction}

An important goal of basic research on quantum materials is to breed new quantum technologies that can address the increasingly complex energy challenges. The past decade has seen a dramatic change in the model for energy production, consumption, and transportation. Due to the explosive growth of wireless technologies and portable devices, there is now increasing effort towards developing microscaled devices that are able to harvest ambient energy into usable electrical energy.
The physical process central to harvesting electromagnetic energy is rectification, which refers to the conversion from an oscillating electromagnetic field to a DC current.  Existing rectifiers operating at radio frequency are mostly based on electrical circuits with diodes, where the built-in electric field in the semiconductor junction sets the direction of the DC current. Such diodes face two fundamental limitations \cite{rectifier,energy}. First, rectification requires a threshold input voltage $V_T=k_B T/e$, known as thermal voltage (about $26\,\text{mV}$ at room temperature). Second, the responsivity is limited by the transition time in diodes (typically order of nanoseconds) and drops at high frequencies.
On the other hand, because of the fast-developing microwatts and nanowatts electronics and next generation wireless networks, energy harvesters of electromagnetic field in the microwave and terahertz (THz) frequency range are in great demand.

High-frequency rectifiers can also be used in sensor and detector technology for the infrared, far infrared and submillimeter bands \cite{detector1,detector2}, which has wide-ranging applications in medicine, biology, climatology, meteorology, telecommunication, astronomy, etc.
However, there is a so-called terahertz gap (0.1 to 10\,THz) between the operating frequencies of electrical diodes and photodiodes.  At frequencies within this range, efficient detection technology remains to be developed.

\begin{figure}
\centering
\includegraphics[width=.85\hsize]{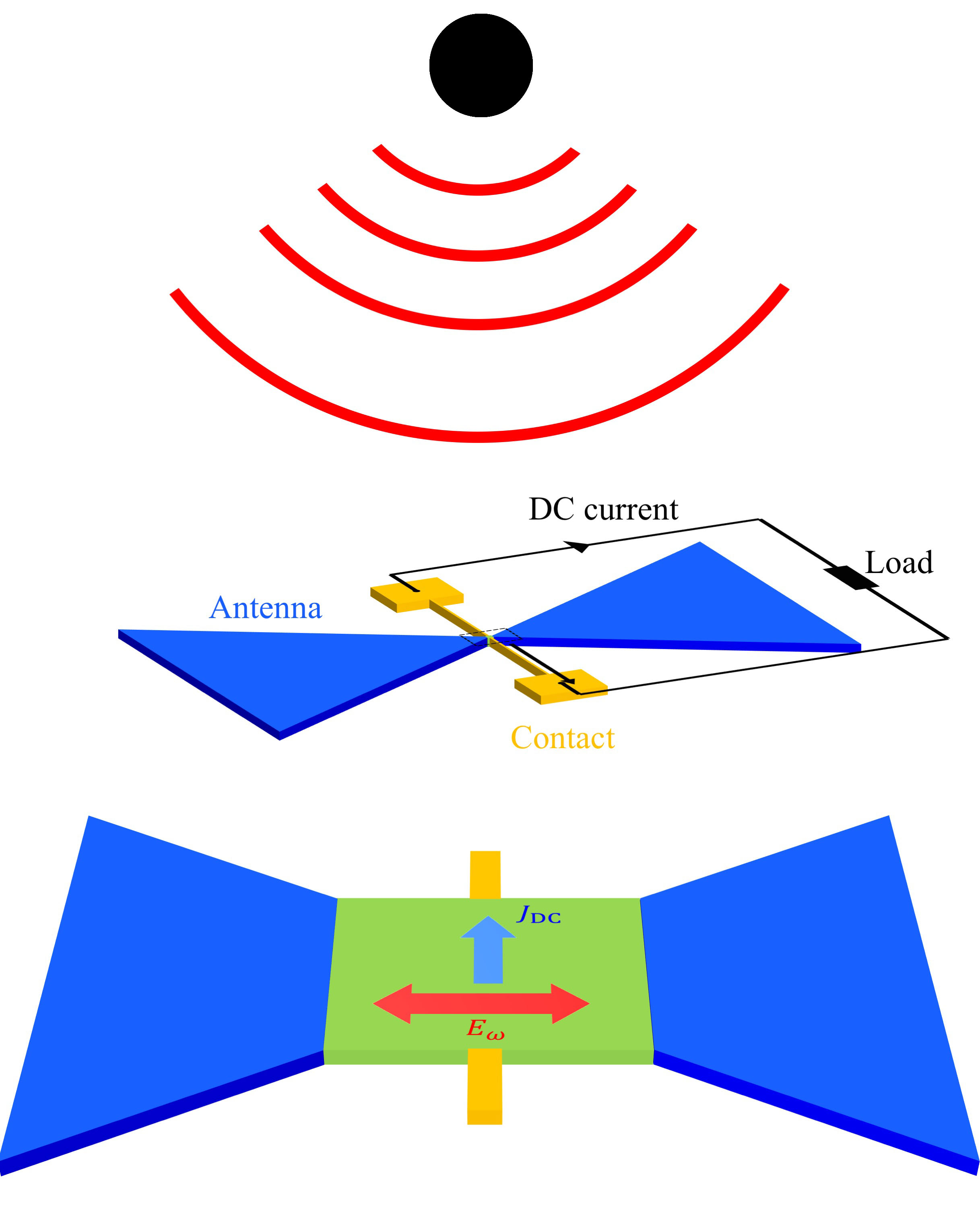}
\caption{
\textbf{Schematic figure of a rectifier based on a two-dimensional material.}
In this setup, we detect the rectified DC current transverse to the incident electric field, which is advantageous in reducing noise.  The antenna is attached to both sides to collect bigger power from radiation and enhance the sensitivity.
}
\label{fig:device}
\end{figure}

Instead of using semiconductor junctions, rectification can be realized as the nonlinear electrical or optical response of noncentrosymmetric crystals (Fig.~\ref{fig:device}). In particular, the second-order nonlinearity $\chi(\omega)$ is an intrinsic material property that characterizes the DC current generated by an external electric field oscillating at frequency $\omega$.
Rectification in a single homogenous material is not limited by the thermal voltage threshold or the transition time innate to a semiconductor junction.
Moreover, nonlinear electrical or optical response of metals and degenerate semiconductors is  much faster than photothermal effects used in bolometers for thermal radiation detection.

However, second-order nonlinearity of most materials is small. While nonlinear optical properties of quantum materials \cite{Boyd,Chemla} such as photocurrent and second-harmonic generation are extensively studied \cite{Weyl2,Weyl3}, much less is known about second-order response at radio, microwave, and infrared frequencies \cite{Genkin,Soref,Leuthold}. In this direction, recent works have predicted intraband photocurrent \cite{Moore,Spivak} and  second-order nonlinear Hall effect due to ``Berry curvature dipole'' \cite{Sodemann} in nonmagnetic materials at zero magnetic field. In particular, the nonlinear Hall effect is predicted to be prominent in materials with titled Dirac or Weyl cones, which are a source of large Berry curvature dipole \cite{Sodemann}. Very recently, this effect was observed for the first time in low-frequency (around 100\,Hz) transport measurements on the two-dimensional (2D) transition metal dichalcogenide (TMD) 1T$'$-WTe$_2$ \cite{dipole, Mak}. The second-order Hall conductivity of bilayer WTe$_2$ is remarkably large \cite{dipole}, in agreement with its large Berry curvature dipole from the tilted Dirac dispersion \cite{Qian}. This and other types of nonlinear response from intraband processes are also being explored in topological insulator surface states, Weyl semimetals, Rashba systems and heavy fermion materials  \cite{Ganichev,Morimoto,Ishizuka,Koenig,nonreciprocal,Zhang,Rostami,Facio,Dzsaber}.
Despite the recent progress, mechanisms for nonlinear electric  and terahertz response of noncentrosymmetric crystals remain to be thoroughly studied.

In this work, we present a systematic theoretical study of intraband second-order response in time-reversal-invariant noncentrosymmetric materials, using a semiclassical Boltzmann equation and taking account of electron Bloch wavefunctions via quantum-mechanical scattering rates.  We identify a new contribution to rectification from skew scattering with nonmagnetic impurities.  
Importantly, such skew scattering arises from the intrinsic chirality of Bloch wavefunctions in momentum space and does {\it not} require spin-orbit coupling.  We show that the contributions to rectification from skew scattering and from Berry curvature dipole scale differently with the impurity concentration, and skew scattering predominates at low impurity concentration. Moreover, skew scattering is allowed in all noncentrosymmetric crystals, whereas Berry curvature dipole requires more strict symmetry conditions.
Based on this new mechanism, we predict large and highly tunable rectification effects in graphene multilayers and heterostructures, as well as 2H-TMD monolayers.

\subsection*{Results}

\subsubsection*{Second-order response}

We study the DC current in a homogeneous material generated at second order by an external electric field $\bm{E}$ of frequency $\omega$.  The second-order response also involves a $2\omega$ component, corresponding to the second-harmonic generation, which we do not focus on in this work.
We write down the second-order DC response as
\begin{equation}
\label{eq:response}
j_a = \chi_{abc} E_b^* E_c,
\end{equation}
where $\chi_{abc}$ is the rank-three tensor for the second-order conductivity, $E_a=E_a(\omega)$ is the (complex) amplitude of the external electric field of frequency $\omega$, and the indices $a,b,c$ label the coordinates.  $\chi_{abc}$ satisfies $\chi_{abc} = \chi_{acb}^*$.  Since the current $j_a$ is odd under inversion while the electric field $E_a$ is even, finite $\chi_{abc}$ is possible only in noncentrosymmetric media.
The second-order response can be decomposed into two parts as
$j_a = \chi_{abc}' E_b^* E_c + i \chi_{abc}'' E_b^* E_c$,
with $\chi_{abc}' = \chi_{acb}'$ and $\chi_{abc}'' = -\chi_{acb}''$.
The former describes the response to a linearly-polarized field and the latter to a circularly-polarized field \cite{Belinicher2}.
It is important to note that in an isotropic medium second-order response to a linearly-polarized field must vanish. In other words, a nonzero $\chi'_{abc}$ requires the presence of crystal anisotropy.

Nonlinear responses are extensively studied in the optical frequency regime \cite{Boyd}.  The classical approach to nonlinear optics considers electrons bound to a nucleus by an anharmonic potential, so that the restoring force to the displacement of electrons becomes nonlinear.  In the quantum-mechanical theory, the nonlinear optical response at frequencies larger than the band gap is usually dominated by electric dipole transitions between different bands.
On the other hand, with energy harvesting and infrared detection in mind,
in this work we consider intraband nonlinear response at frequencies below the interband transition threshold.

\subsubsection*{Semiclassical Boltzmann analysis}

We analyze the second-order electrical transport using the semiclassical Boltzmann theory \cite{Belinicher2}.
For a homogeneous system, the Boltzmann equation is given by $(\hbar=1)$
\begin{equation}
\label{eq:boltzmann}
\frac{\partial f}{\partial t} + \dot{\bm{k}}\cdot\frac{\partial f}{\partial \bm{k}} = -\mathcal{C}[f],
\end{equation}
where $f(\bm{k},\epsilon)$ is the distribution function at energy $\epsilon$ and $\mathcal{C}[f]$ denotes the collision integral.  The time derivative of the wavevector is equal to the force felt by an electron $\dot{\bm{k}}=-e\bm{E}$ under the external electric field $\bm{E}$.
The distribution function $f$ deviates from the equilibrium Fermi--Dirac distribution when electrons are accelerated by the external field. A nonequilibrium steady state is obtained by the balance of the acceleration and the relaxation due to scattering, described by the collision integral $\mathcal{C}[f]$, whose explicit form is shown later in a general form.  It includes the scattering rate $w_{\bm{k}'\bm{k}}$, the probability of the transition per unit time from a Bloch state with a wavevector $\bm{k}$ to that with $\bm{k}'$.
We will consider scattering by impurities.  The impurity density should be sufficiently lower than the electron density for the semiclassical Boltzmann analysis to be valid; see Materials and Methods for further discussions. 

To obtain a nonzero second-order conductivity $\chi_{abc}$, we need to capture the effect of inversion breaking.
For time-reversal-invariant systems, the energy dispersion $\epsilon_{\bm{k}}$ and the band velocity $\bm{v}_0(\bm{k}) = \nabla_{\bm{k}}\epsilon_{\bm{k}}$ satisfy the conditions $\epsilon_{\bm{k}}=\epsilon_{-\bm{k}}$ and $\bm{v}_0(\bm{k}) = -\bm{v}_0(-\bm{k})$, which hold with or without inversion symmetry: that is, the absence of inversion is encoded in the wavefunction but not in the energy dispersion.

One consequence of inversion asymmetry is manifested in a scattering process, as the transition rate depends not only on the dispersion but also on the Bloch wavefunction $\psi_{\bm{k}}$. For and only for noncentrosymmetric crystals, the transition rate from $\bm{k}$ to $\bm{k}'$ and from $-\bm{k}$ to $-\bm{k}'$ can be different: $w_{\bm{k}'\bm{k}} \neq w_{-\bm{k}',-\bm{k}}$. A scattering process with such asymmetry is referred to as skew scattering.  We will see that skew scattering is a source of the second-order DC response, i.e., rectification.
The strength of skew scattering  is characterized by $w_{\bm{k}'\bm{k}}^{(A)}=(w_{\bm{k}'\bm{k}}-w_{-\bm{k}',-\bm{k}})/2$, whereas the symmetric component $w_{\bm{k}'\bm{k}}^{(S)}=(w_{\bm{k}'\bm{k}}+w_{-\bm{k}',-\bm{k}})/2$ does not rely on inversion breaking.  The presence of time-reversal symmetry guarantees $w_{\bm{k}'\bm{k}} = w_{-\bm{k},-\bm{k}'}$, which allows us to write $w_{\bm{k}'\bm{k}}^{(S,A)}=(w_{\bm{k}'\bm{k}}\pm w_{\bm{k}\bm{k}'})/2$.

The transition rate $w_{\bm{k}'\bm{k}}$ is usually calculated to the lowest order in scattering potential by Fermi's golden rule.  For elastic impurity scattering, it is given by $w_{\bm{k}'\bm{k}}^{(S)} = 2\pi \langle |V_{\bm{k}'\bm{k}}|^2 \rangle \delta(\epsilon_{\bm{k}'}-\epsilon_{\bm{k}})$, where $V_{\bm{k}'\bm{k}}$ is the matrix element of a single scatterer and $\langle \ \rangle$ denotes the impurity average.  However, this formula is symmetric under the exchange $\bm{k}\leftrightarrow\bm{k}'$ and does not capture skew scattering.
The latter arises at the next-leading order from the interference of a direct transition $\bm{k}\to\bm{k}'$ and an indirect process $\bm{k}\to\bm{k}'$ via an intermediate state $\bm{q}$ \cite{Luttinger1}:
\begin{equation}
w_{\bm{k}'\bm{k}}^{(A)} = -(2\pi)^2 \int_{\bm{q}} \operatorname{Im} \langle V_{\bm{k}'\bm{q}} V_{\bm{q}\bm{k}} V_{\bm{k}\bm{k}'} \rangle \delta (\epsilon_{\bm{k}}-\epsilon_{\bm{k}'}) \delta (\epsilon_{\bm{k}'}-\epsilon_{\bm{q}}), \label{wA}
\end{equation}
which indeed satisfies $w_{\bm{k}'\bm{k}}^{(A)} = -w_{\bm{k}\bm{k}'}^{(A)}$. Here we use the notation $\int_{\bm{q}}=\int d^dq/(2\pi)^d$ ($d$ is the spatial dimension).

\begin{figure*}
\centering
\includegraphics[width=0.7\hsize]{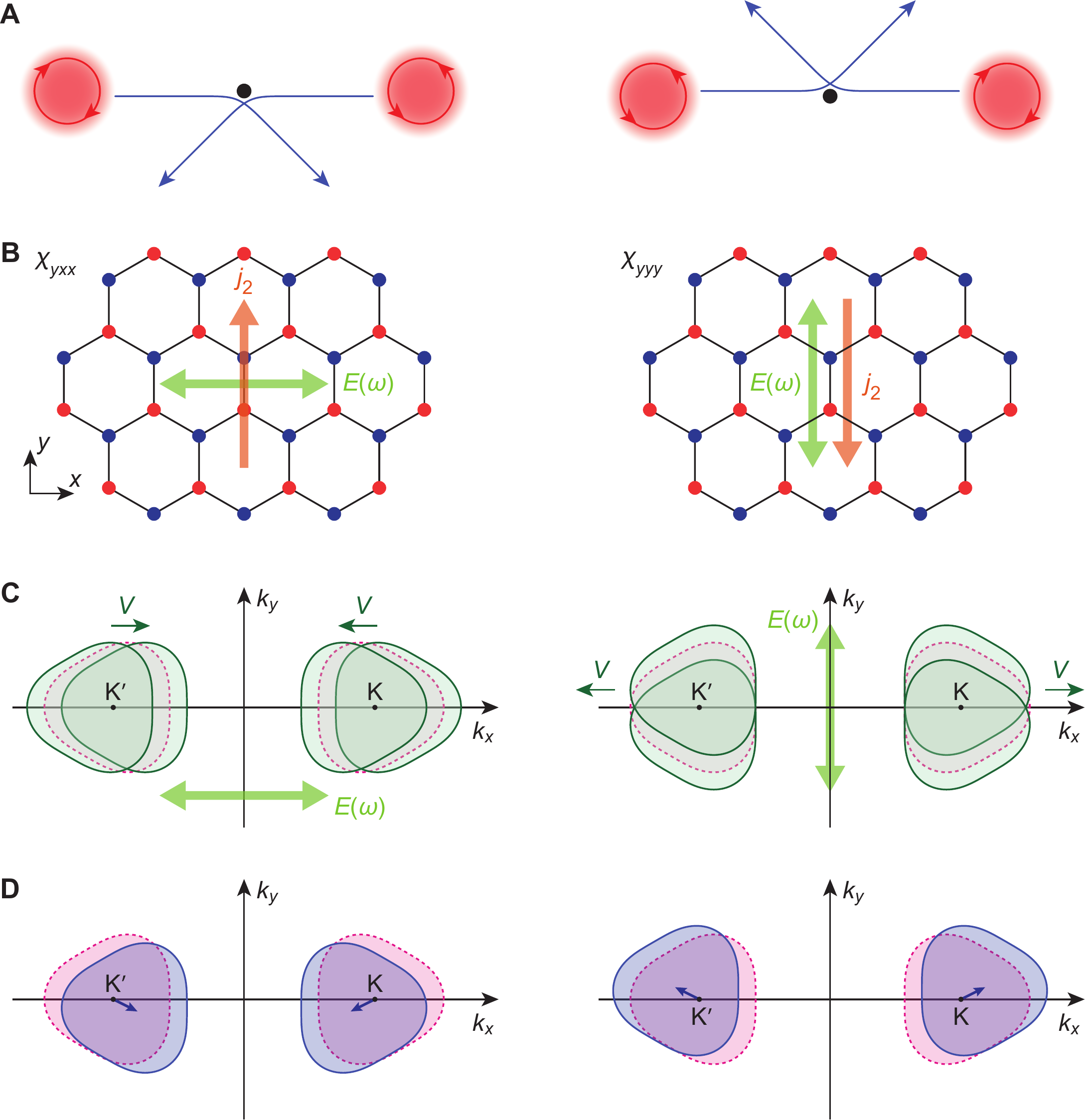}
\caption{
\textbf{Second-order response by skew scattering on a honeycomb lattice.}
(\textbf{A}) Schematics of skew scattering.  When a self-rotating wave packet (red) is scattered by an inversion-symmetric potential (black), its motion is deflected like the Magnus effect.  Two wave packets moving in the opposite directions produces zero net current.  However, if the two self-rotate in different directions, skew scattering produces net current in the perpendicular direction.
(\textbf{B}) Electric field and rectified current on a honeycomb lattice.  The left and right panels in (A) to (D) correspond to $\chi_{yxx}$ and $\chi_{yyy}$, respectively.
(\textbf{C}) Fermi surface displacement at frequency $\omega$ (green).  The oscillating electric field $\bm{E}(\omega)$ forces the Fermi surface to swing back and forth from its equilibrium position (red).  Owing to the Fermi surface anisotropy, each valley yields finite velocity $\bm{V}$ along the $k_x$ direction after time averaging.  This velocity is canceled with the two valleys, and there is no DC current generated as a linear response.
(\textbf{D}) Stationary Fermi surface displacement.  The electric field and skew scattering produce the stationary Fermi surface displacement (blue) from the equilibrium state (red) as a second-order response.  Finite rectified current is observed when the contributions from the two valleys do not cancel.
}
\label{fig:geometry}
\end{figure*}

It is well known that skew scattering provides an extrinsic contribution to the anomalous Hall effect \cite{AHE} and the spin Hall effect \cite{SHE}. These are linear response phenomena in systems with broken time-reversal and spin-rotational symmetries, respectively. We emphasize that skew scattering exists and contributes to second-order response even in materials with both time-reversal and spin-rotational symmetry. With regard to its microscopic origin, skew scattering can be caused by nonmagnetic impurities when the single-impurity potential $V(\bm{r})$ is inversion asymmetric, such as a dipole or octupole potential \cite{Belinicher2}. Here we show that even for a symmetric single-impurity potential $V(\bm{r})=V(-\bm{r})$ such as a delta function potential, skew scattering can also appear. In such case, skew scattering is due to the inherent chirality of the electron wavefunction in a noncentrosymmetric crystal.
In the simplest case of a delta function impurity potential, we have $V_{\bm{k}'\bm{k}} \propto \langle u_{\bm{k}'} | u_{\bm{k}} \rangle$ where $| u_{\bm{k}} \rangle$ is the Bloch wavefunction within a unit cell. Then the expression for $w^{(A)}$ given in Eq.~\eqref{wA} involves the wavefunction overlap at different $\bm{k}$ points on the Fermi surface, similar to the Berry phase formula \cite{Vanderbilt}, 
and also skew scattering is known to be related to Berry curvature in a certain limit \cite{Nagaosa}. 
Importantly, while there is no net chirality in nonmagnetic systems, the Bloch wavefunction $| u_{\bm{k}} \rangle$ of an electron at finite momentum is allowed to be complex and carry $\bm k$-dependent orbital magnetization, spin polarization, or Berry curvature. In other words, mobile electrons in inversion-breaking nonmagnetic crystals can exhibit momentum-dependent chirality that is opposite at $\bm k$ and $-\bm k$.
Then, skew scattering occurs due to electron chirality in a similar way as the classical Magnus effect, where a spinning object is deflected when moving in a viscous medium.  The spinning motion is associated with the self-rotation of a wave packet \cite{Chang} and the viscosity results from scattering with impurities; see Fig.~\ref{fig:geometry}A.

To obtain the second-order conductivity, the distribution function $f(\bm{k})$ should be calculated to second order in the external electric field $\bm{E}(\omega)$.
Moreover, since skew scattering $w^{(A)}$ is parametrically smaller than $w^{(S)}$, we expand the distribution function up to first order in $w^{(A)}$.
By solving the Boltzmann equation self-consistently at each order of $\bm{E}$ and $w^{(A)}$, we obtain the second-order DC current response $\bm{j}=-e\int_{\bm{k}} \bm{v}_0(\bm{k}) f(\bm{k})$ due to skew scattering in time-reversal-invariant systems.
The second-order conductivity $\chi_{abc}$ can be formally expressed in terms of various scattering times $\tau_n, \tau'_n$:
\begin{align}
\label{eq:chi}
\chi_{abc} = &-e^3 \bigg[ \tau'_2 \tau'_{1}(\omega) \tau_{1} (\omega) \int_{\bm{k}} v_{0,a} \partial_{k_b} \int_{\bm{k}'} w_{\bm{k}\bm{k}'}^{(A)} \partial_{k'_c} f_0(\bm{k}') \nonumber\\
&+ \tau'_2 \tau_2 \tau_{1}(\omega)
\int_{\bm{k}} v_{0,a} \int_{\bm{k}'} w_{\bm{k}\bm{k}'}^{(A)} \partial_{k'_b} \partial_{k'_c} f_0(\bm{k}')
\bigg] \nonumber\\
& + (b \leftrightarrow c)^* ,
\end{align}
where $\tau_n(\omega)$ is a shorthand for $\tau_n(\omega) = \tau_n/(1-i\omega\tau_n)$ and likewise for $\tau'_n(\omega)$. The scattering times $\tau_n, \tau'_n$ with $n=1,2$ are associated with the dominant symmetric scattering $w^{(S)}$. They determine the steady-state distribution function up to second order in the electric field, when skew scattering is neglected. Detailed descriptions of a self-consistent solution of the Boltzmann equation and discussions about Joule heating can be found in Supplementary Materials.

A rough order-of-magnitude estimate of $\chi$ is obtained from Eq.~(\ref{eq:chi}) by using two scattering times $\tau$ and $\tilde{\tau}$ for the symmetric and skew scattering, respectively, leading to the second-order DC current of a metal or degenerate semiconductor under a linearly-polarized electric field: 
\begin{equation}
\label{eq:chi_approx}
j_{2}  \sim
en v_F \cdot \left( \dfrac{e E \Delta t}{p_F} \right)^2 \cdot \dfrac{\tau}{\tilde{\tau}}
\end{equation}
with $\Delta t=\tau$ for $\omega\tau \ll 1$ and $\Delta t = 1/\omega$ for $\omega\tau \gg 1$.
Here $v_F$ is the Fermi velocity, $p_F=\hbar k_F$ is the Fermi momentum, and $n$ is the electron density.

This estimate provides a heuristic understanding of the second-order response.
In the low-frequency limit $\omega \rightarrow 0$, the ratio of the second-order current $j_2$ and the linear response current $j_1=\sigma_\omega E$ is a product of two dimensionless quantities $eE \tau /p_F$ and $\tau/\tilde{\tau}$. The first term is the change of electron's momentum under the external field during the mean free time, divided by the Fermi momentum. The second term $\tau/\tilde{\tau}$ characterizes the strength of skew scattering which is responsible for second-order electrical response, relative to the symmetric scattering.

The short-circuit current responsivity $\mathfrak{R}_I$ is a metric for a rectifier, defined as the ratio of the generated DC current to the power dissipation.  With a sample dimension $L^2$, the current responsivity is given by $\mathfrak{R}_I \equiv j_2L/(j_1EL^2)$.
Similarly, the voltage responsivity $\mathfrak{R}_V$ is defined for the generated DC voltage as $\mathfrak{R}_V \equiv (j_2L/\sigma_0)/(j_1EL^2)$.
Both $\mathfrak{R}_I$ and $\mathfrak{R}_V$ are independent of the magnitude of the external electric field. Using the linear response conductivity $\sigma_\omega \sim env_F \cdot (e\lambda_F\tau/\hbar) \operatorname{Re}[(1-i\omega\tau)^{-1}]$ obtained from the same Boltzmann equation, we have
\begin{equation}
\label{eq:responsivity}
\begin{gathered}
\mathfrak{R}_I = \frac{1}{L} \frac{\chi}{\sigma_\omega} = \frac{\eta_I}{L}\frac{\tau}{\tilde{\tau}},
\quad \eta_I \sim \frac{e\tau}{p_F}, \\
\mathfrak{R}_V = \frac{1}{L} \frac{\chi}{\sigma_\omega \sigma_0} = \frac{\eta_V}{L}\frac{\tau}{\tilde{\tau}} ,
\quad \eta_V \sim \frac{1}{env_F},
\end{gathered}
\end{equation}
Here we define the reduced current and voltage responsivities $\eta_I$ and $\eta_V$, respectively, which are independent of the sample size and the ratio $\tau/\tilde{\tau}$.
The approximate relations for $\eta_I$ and $\eta_V$ hold in both low- and high-frequency limits.
We also consider the ratio of the generated DC power and input power $\eta_P \equiv (j_2^2/\sigma_0)/(\sigma_\omega E^2)$, which characterizes the power conversion efficiency (assuming the load resistance and internal resistance are comparable). We find
\begin{eqnarray}
\eta_P = \frac{\chi^2 E^2}{\sigma_0 \sigma_\omega}
\propto \left( \frac{e E \Delta t}{p_F}\right)^2 \cdot \left(\frac{\tau}{\tilde{\tau}} \right)^2.
\end{eqnarray}
From these figure of merits, it is clear that a material with low carrier density $n$ or small Fermi momentum $p_F$ is desirable for efficient rectification.
Moreover, to achieve high current responsivity a long mean free time is preferred.
In this respect, high-mobility semiconductors and semimetals are promising as rectifiers for infrared detection.
In the following, we analyze graphene systems and estimate their efficiency of rectification.

\subsubsection*{Graphene multilayers and transition metal dichalcogenides}

Pristine graphene has high mobility and low carrier density, but it is centrosymmetric.  Nevertheless, as a van der Waals material, we can easily assemble multilayer stacks to break inversion.  Realizations of noncentrosymmetric structures include monolayer graphene on a hexagonal boron nitride (hBN) substrate \cite{Louie,LeRoy,Pablo1,Dean1,Wallece}, electrically-biased bilayer graphene \cite{McCann,graphene_review1,graphene_review2}, and multilayer graphene such as ABA-stacked trilayer graphene \cite{Wu,Young}.
Anisotropy of energy dispersions, also required for second-order response, naturally arises from trigonal crystal structures.

We now show how skew scattering arises from the chirality of quantum wavefunctions and contributes to the second-order conductivity in graphene heterostructures with gaps induced by inversion symmetry breaking. The $k \cdot p$ Hamiltonian we consider here is
\begin{align}
H_s(\bm{k}) &=
\begin{pmatrix}
\Delta & svk_{-s} -\lambda k_s^2 \\
svk_s - \lambda k_{-s}^2 & -\Delta
\end{pmatrix},
\label{eq:hamiltonian}
\end{align}
where $s=\pm1$ denotes two valleys at $K$ and $K'$, respectively, and we define $k_\pm = k_x\pm ik_y$. This Hamiltonian has a band gap of $2\Delta$ in the energy spectrum and describes two graphene systems: (i) monolayer graphene on hBN, where the spatially varying atomic registries break the carbon sublattice symmetry and opens up gaps at Dirac points \cite{Louie,LeRoy, Pablo1,Dean1,Wallece} and (ii) bilayer graphene with an out-of-plane electric field that breaks the layer symmetry and opens up gaps \cite{McCann,graphene_review1,graphene_review2}.
The difference between the monolayer and bilayer cases lies in the relative strength of $v$ and $\lambda$. For the monolayer (bilayer) case, $\lambda k_F^2 \ll vk_F$ ($vk_F\ll \lambda k_F^2$) is responsible for the trigonal warping of the linear (quadratic) energy dispersion.

The Bloch wavefunction of $H_s(\bm{k})$ has two components associated with sublattice (layer) degrees of freedom in monolayer (bilayer) graphene. When the band gap is present ($\Delta \neq 0$), the wavefunction in each valley is chiral and carries finite Berry curvature, leading to skew scattering from nonmagnetic impurities. The chirality and Berry curvature are opposite for valley $K$ and $K'$ due to time reversal symmetry. Note that because of the three-fold rotation symmetry of graphene, Berry curvature dipole vanishes \cite{Sodemann}, leaving skew scattering as the only mechanism for rectification.

We calculate the second-order conductivity from Eq.~\eqref{eq:chi}, and find nonvanishing elements $\chi_{xxy} = \chi_{xyx} = \chi_{yxx} = -\chi_{yyy} \equiv \chi$, consistent with the point group symmetry.  The relation among the electric field, induced current, and underlying crystalline lattice is depicted in Fig.~\ref{fig:geometry}B, along with the schematic pictures of the Fermi surface displacement (Fig.~\ref{fig:geometry}, C and D).
The function $\chi$ has the form $\chi = e^3 v_F (\tau^3/\tilde{\tau}) \zeta(\epsilon_F,\omega)$, with the dimensionless function $\zeta$ given in the Supplementary Materials.
The ratio $\tau/\tilde{\tau}$ is proportional to the gap $\Delta$ and the Fermi surface trigonal warping. The former describes the effect of inversion breaking on the electronic structure, while the latter is responsible for Fermi surface asymmetry within a valley $E_s(\bm{k})\neq E_s(-\bm{k})$. Both ingredients are necessary for finite second-order response.

\begin{figure*}
\centering
\includegraphics[width=0.7\hsize]{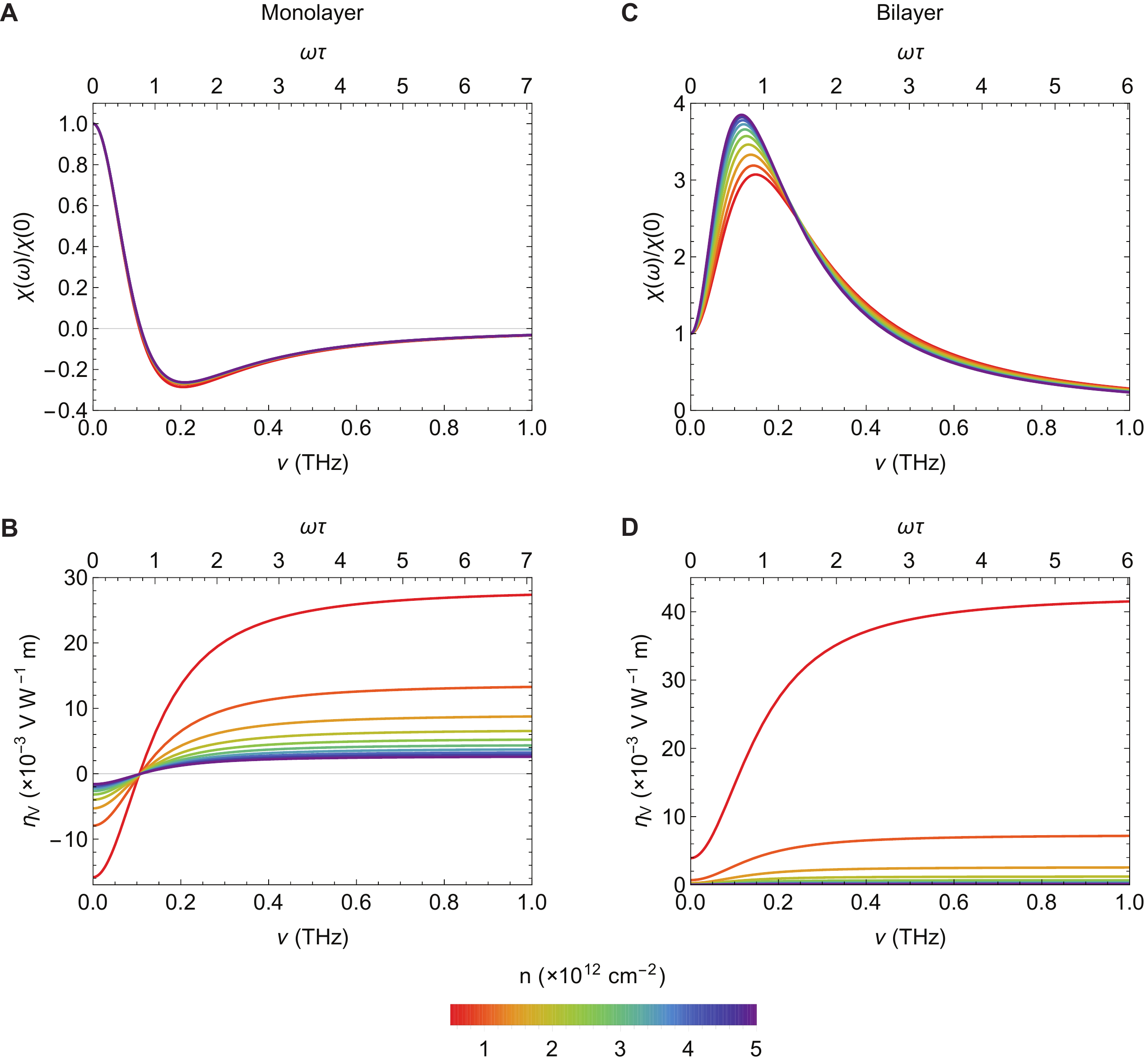}
\caption{
\textbf{Second-order response for graphene heterostructures and multilayers.}
(\textbf{A} and \textbf{B}) Second-order DC response for monolayer and (\textbf{C} and \textbf{D}) bilayer graphene.  The second-order conductivity $\chi$ is shown in (A) and (C) and the reduced voltage responsivity $\eta_V$ in (B) and (D).
The carrier density $n$ is changed from $0.5\times 10^{12}\,\text{cm}^{-2}$ (red) to $5\times 10^{12}\,\text{cm}^{-2}$ (purple).
We use the values $v=0.94\times 10^6\,\text{m/s}$, $\Delta=15\,\text{meV}$, and $\tau=1.13\,\text{ps}$ for the monolayer case; $\Delta=50\,\text{meV}$ and $\tau=0.96\,\text{ps}$ for the bilayer case, from transport, infrared spectroscopy, and scanning tunneling microscopy/spectroscopy measurements \cite{Dean1,Wang,Dean2,Xue,Lu,Schmitz}.
For the bilayer case $\lambda=(2m)^{-1}$ is determined by the effective mass $m\approx 0.033m_e$ ($m_e$: electron mass) and $v\approx 1\times 10^5\,\text{m/s}$ \cite{bilayer1,bilayer2}.
See also the Supplementary Materials for details.
}
\label{fig:efficiency}
\end{figure*}

The second-order conductivity $\chi$ and the reduced voltage responsivity $\eta_V$ for both monolayer and bilayer gapped graphene are shown in Fig.~\ref{fig:efficiency} as the functions of frequency $\nu=\omega/(2\pi)$. The frequency dependence of the second-order conductivity is qualitatively understood with Eq.~\eqref{eq:chi_approx}.  We can further simplify the relation in two dimensions for $n\lambda_F^2 = \text{const.}$ The second-order conductivity then behaves as $\chi \sim e^3v_F\tau^2/\hbar^2 \cdot (\tau/\tilde{\tau})$ for $\omega\tau\ll 1$ and $\chi \sim e^3v_F/(\hbar\omega)^2 \cdot (\tau/\tilde{\tau})$ for $\omega\tau \gg 1$.

The responsivity is affected by the linear conductivity $\sigma$ as well as the second-order conductivity $\chi$.
At high frequencies with the carrier density fixed, we observe a decrease of the rectified current while the energy dissipation by the linear response also decreases.
Since both $\sigma$ and $\chi$ decrease as $\omega^{-2}$, the responsivity saturates in the high frequency limit.
With a sample size of $10\,\mu\text{m}$, the carrier density $n=0.5\times 10^{12}\,\text{cm}^{-2}$, and the ratio $\tau/\tilde{\tau}=0.01$, from reasonable estimates of mean free time and impurity concentration
we find the saturated voltage responsivity $\mathfrak{R}_V \sim 30\,\text{V/W}$ for monolayer and $40\,\text{V/W}$ for bilayer (see also the Supplementary Materials for the frequency dependence).
As expected from  the qualitative estimate Eq.~\eqref{eq:responsivity}, high responsivity is realized with low carrier density.  We note that the semiclassical approach is valid when the mean free path  is much longer than the Fermi wavelength.

Both monolayer and bilayer graphene show broadband response.  The saturation of responsivities lasts up to the onset frequency of an interband transition (not included in our study).
For example, at the lowest density of $n = 0.5\times 10^{12}\,\text{cm}^{-2}$ shown in Fig.~\ref{fig:efficiency}, the Fermi energy for the monolayer case  is $\epsilon_F \approx 80\,\text{meV}$, so that due to the Pauli blocking the interband transition does not occur at frequencies below $2\epsilon_F \sim 40\,\text{THz}$.  The interband threshold frequency also increases with the band gap, which is tunable in bilayer graphene by the displacement field.

Finite temperature affects the intraband second-order response and responsivity through thermal smearing of the distribution function and the change of scattering times. Here thermal smearing does not change our result much as the Fermi energy is much higher than room temperature. Since it utilizes the material's intrinsic nonlinearity, rectification from the intraband process considered here does not suffer from the noise associated with thermally excited electron-hole pairs in photodiodes.

Our analysis also applies to 2H-TMD monolayers, where transition metal and chalcogen atoms form trigonal crystal structures. Similar to gapped graphene, their band structures can also be described as massive Dirac fermions with trigonal warping \cite{TMD_warping}, but with much larger band gaps \cite{TMD_review,TMD_review2}. Hence they should exhibit large intraband second-order response as well.  Topological insulator surface states also shows the effect (see Supplementary Materials). In addition to 2D systems, 3D bulk materials without inversion are also worth considering.  As discussed, we expect large responsivities in  low carrier densities.  From this respect, inversion-breaking Weyl semimetals are promising candidates.

\subsection*{Discussion}

So far we have considered the skew scattering contribution to the second-order DC response.  Besides, the Berry curvature $\bm{\Omega}$ contributes to the distribution function through the collision integral and the electron's velocity as the anomalous and side-jump velocities, which modifies the second-order conductivity.  The Berry curvature is odd (even) under time reversal (inversion), and thus it is allowed to exist in time-reversal-invariant noncentrosymmetric materials.  
A similar Boltzmann transport analysis for linear response is reported in the context of the anomalous Hall effect without time-reversal symmetry \cite{AHE}; see Materials and Methods, and Supplementary Materials for detail.  Finite Berry curvature enforces a coordinate shift $\delta\bm{r}_{\bm{k}\bm{k}'}$ after scattering, giving a displacement of the center of an electron wave packet.  It is accompanied by a potential energy shift $\delta\epsilon_{\bm{k}'\bm{k}}$ in the presence of the external field, which results in the collision integral
\begin{equation}
\mathcal{C}[f] = \int_{\bm{k}'} \left[ w_{\bm{k}'\bm{k}} f(\bm{k},\epsilon_{\bm{k}}) -w_{\bm{k}\bm{k}'} f(\bm{k}',\epsilon_{\bm{k}}+\delta\epsilon_{\bm{k}'\bm{k}}) \right].
\end{equation}

One may decompose the distribution function in nonequilibrium as $f=f_0 + f^\text{scatt} + f^\text{adist}$, where $f_0$ is the equilibrium distribution function, $f^\text{scatt}$ describes the effect of scattering to the distribution (with $\delta\epsilon_{\bm{k}'\bm{k}}=0$), and $f^\text{adist}$ is the anomalous distribution due to finite $\delta\epsilon_{\bm{k}'\bm{k}}$.
Because of the the different origins, they have distinct dependence on the scattering time $\tau$.  For low frequencies with $\omega\tau \ll 1$, $\tau$ dependence of each term is easily estimated by a simple power counting. Noting that the scattering rate $w$ amounts to $\tau^{-1}$, we can expand the distribution function in powers of the electric field $E$, with coefficients depending on $\tau$:
\begin{equation}
\begin{aligned}
f^\text{scatt}&=\sum_{n\geq 1} f_n^\text{scatt},& f_n^\text{scatt} &\propto \tau^n E^n, \\
f^\text{adist}&=\sum_{n \geq 1} f_n^\text{adist},& f_n^\text{adist} &\propto \tau^{n-1} E^n,
\end{aligned}
\end{equation}
in the weak impurity limit.

The electric current density is obtained by $\bm{j} = -e \int_{\bm{k}} \bm{v}(\bm{k}) f(\bm{k})$, where $\bm{v}(\bm{k})$ is the electron's velocity,
\begin{eqnarray}
\bm{v} &=& \frac{\partial\epsilon_{\bm{k}}}{\partial\bm{k}} - \dot{\bm{k}} \times \bm{\Omega} + \int_{\bm{k}'} w_{\bm{k}'\bm{k}} \delta\bm{r}_{\bm{k}'\bm{k}} \nonumber \\
&\equiv& \bm{v}_0 + \bm{v}_\text{av} + \bm{v}_\text{sj}.
\end{eqnarray}
Besides the velocity given by the band dispersion $\bm{v}_0$, $\bm{v}$ contains two additional terms: The anomalous velocity $\bm{v}_\text{av}$ is associated with the Berry curvature and the last term $\bm{v}_\text{sj}$ describes the side-jump contribution, which arises from a combined effect of scattering and Berry curvature.
Now, we can see several contributions to the second-order conductivity $\chi$ from combinations in the product of the out-of-equilibrium distribution function $(f^\text{scatt} + f^\text{adist})$ and the electron velocity $(\bm{v}_0 + \bm{v}_\text{av} + \bm{v}_\text{sj})$.

The dominant contribution in the weak impurity limit $(\tau\to\infty)$ emerges from the skew scattering with $\bm{v}_0 f^\text{scatt}_2$, resulting in $\chi\propto\tau^2$.
The Berry curvature dipole contribution found in Ref.~\cite{Sodemann} corresponds to $\bm{v}_\text{av}f^\text{scatt}_1$, which amounts to $\chi\propto\tau$.
It shows the frequency dependence as
$\chi'\propto e^3 \tau D$, $\chi'' \approx 0$ for low frequencies $(\omega\tau\ll 1)$ and $\chi' \propto e^3 D/(\omega^2 \tau)$, $\chi''\propto e^3D/\omega$ for high frequencies $(\omega\tau\gg1)$, where $D$ is the Berry curvature dipole.  It is a quantity with the dimension of length in two dimensions.
The Berry curvature dipole and skew scattering have the same frequency dependence although they are distinct in the $\tau$ dependence.

Inversion breaking results in finite skew scattering $w^{(A)}$.  Along with the accompanying anisotropic Fermi surface, the second-order electronic response induced by skew scattering persists in any noncentrosymmetric materials.  This is in marked contrast to the Berry curvature dipole mechanism, which imposes more symmetry constraints in addition to inversion breaking.  Some symmetry classes without inversion, e.g., those containing three-fold rotation axis, do not show nonlinear response from a Berry curvature dipole \cite{Sodemann}.

In summary, we have performed a systematic study of second-order electrical response due to intraband process and have identified the skew scattering mechanism as the dominant contribution in the weak impurity limit.  This mechanism is responsible for current rectification and opens a new way to low-power energy harvesters and terahertz detection.

\subsection*{Materials and Methods}

We employ the semiclassical Boltzmann transport theory to analyze the second-order response in noncentrosymmetric materials.  
The analysis is to some extent in parallel with that for the anomalous Hall effect \cite{Belinicher2,AHE}, but there is a fundamental difference in symmetry: Inversion should be broken for finite second-order response whereas time-reversal symmetry is broken for the anomalous Hall effect.

The total Hamiltonian is given by
\begin{equation}
H = H_0 + V + U,
\end{equation}
where $H_0$ is the Hamiltonian for electrons in the clean limit with the eigenvalue $\epsilon_{\bm{k}}$, $V$ describes electron scattering, and $U$ includes the external electric field, which drives the system into nonequilibrium.  In a semiclassical description, $U$ results in the force acting on an electron's wave packet: $\nabla_{\bm{r}} U=-\bm{F}$.  In the following, we consider an external electric field of frequency $\omega$, namely, $\bm{F} = -e\bm{E}(\omega)$.

A formally rigorous analysis requires a quantum-mechanical calculation of density matrices, which usually requires considerable effort.  
The semiclassical analysis is simpler and offers intuitive understanding; the downside is that it has limitations owing to the uncertainty principle.  The semiclassical description relies on wave packets of Bloch states.  A wave packet is localized both in the real space and the momentum space so that the mean position $\bm{r}$ and momentum $\bm{k}$ can be designated.  However, the uncertainty principle imposes the limitation $\Delta r \Delta k \gtrsim 1$. 
We require position and momentum resolutions set by the mean free path $\ell$ and the Fermi wavevector $k_F$; i.e., $k_F \ell \gg 1$ is necessary for the semiclassical analysis.  In the following, we deal with electron scattering by impurities.  Since $\ell$ and $k_F$ depends on the impurity and electron densities, respectively, the condition is satisfied when the impurity density is sufficiently lower than the electron density. 

The semiclassical Boltzmann equation for the distribution function $f$ includes the scattering rate $w_{\bm{k}'\bm{k}}$ and the energy shift $\delta\epsilon_{\bm{k}'\bm{k}}$.  The scattering rate is given by 
\begin{equation}
w_{\bm{k}'\bm{k}} = 2\pi |T_{\bm{k}'\bm{k}}| \delta (\epsilon_{\bm{k}'} - \epsilon_{\bm{k}}),
\end{equation}
with the scattering $T$ matrix $T_{\bm{k}'\bm{k}}$.
The energy shift appears for a scattering process in the presence the Berry curvature $\bm{\Omega}$.  
It is defined from the Berry connection $A_a = i \Braket{u_{\bm{k}}|\partial_{k_a} u_{\bm{k}}}$ as $\Omega_a = \varepsilon_{abc} \partial_{k_b} A_c$, where $\ket{u_{\bm{k}}}$ is the lattice periodic part of the Bloch function at $\bm{k}$ and $\varepsilon_{abc}$ is the Levi-Civita symbol.
The Berry curvature transforms under inversion $\mathcal{P}$ and time reversal $\mathcal{T}$ as
\begin{gather}
\mathcal{P}: \Omega_a(\bm{k}) \to \Omega_a(-\bm{k}), \quad
\mathcal{T}: \Omega_a(\bm{k}) \to -\Omega_a(-\bm{k}).
\end{gather}
It is deduced that the Berry curvature vanishes everywhere in the Brillouin zone when $\mathcal{PT}$ symmetry exists.
Finite Berry curvature causes the coordinate shift $\delta\bm{r}_{\bm{k}'\bm{k}}$, which describes the displacement of the center of a wave packet after a scattering process $\bm{k}\to\bm{k}'$.  
For a weak scattering with a small momentum change, the coordinate shift can be approximated as $\delta\bm{r}_{\bm{k}'\bm{k}} \approx (\bm{k}'-\bm{k}) \times \bm{\Omega}(\bm{k})$.
The energy shift $\delta\epsilon_{\bm{k}'\bm{k}}$ is given by using the coordinate shift $\delta\bm{r}_{\bm{k}'\bm{k}}$ as
\begin{equation}
\delta\epsilon_{\bm{k}'\bm{k}} = -\bm{F} \cdot \delta\bm{r}_{\bm{k}'\bm{k}}.
\end{equation}
It is worth noting that the coordinate shift is independent of the scattering time $\tau$ even though it is related to scattering as it describes the displacement after a single impurity scattering.
Further details of the analysis are presented in the Supplementary Materials.


\subsection*{Acknowledgments}

We thank Q. Ma, M. Solja\v{c}i\'c, B. Halperin, and L. Liu for helpful discussions. 
\textbf{Funding:} The work was supported by the US Army Research Laboratory and the US Army Research Office through the Institute for Soldier Nanotechnologies. L.F was supported in part by a Simons Investigator Award from the Simons Foundation.
\textbf{Author contributions:} H.I. and L.F. conceived the idea.  H.I. performed the theoretical calculations.  All authors discussed the results and wrote the manuscript. 
\textbf{Competing interests:} The authors are part of inventors on a patent application related to this work filed by the MIT (application no.~62/779,025).  They declare that they have no other competing interests. 
\textbf{Data and materials availability:} All data needed to evaluate the conclusions in the paper are present in the paper and/or the Supplementary Materials.  Additional data related to this paper may be requested from the authors.


\clearpage

\onecolumngrid
\noindent

\begin{center}
{\large\bf Supplementary Materials}
\end{center}

\setcounter{section}{0}
\setcounter{equation}{0}
\setcounter{figure}{0}
\def\thesection{S\arabic{section}}
\def\thesubsection{S\arabic{section}.\arabic{subsection}}
\def\theequation{S\arabic{equation}}
\def\thefigure{S\arabic{figure}}



\section{Semiclassical Boltzmann theory}

We describe in detail the analysis of the semiclassical Boltzmann transport theory for the calculation of second-order response in noncentrosymmetric materials.
We deal with spatially homogeneous systems and hence the electron distribution function $f(\bm{k},\epsilon)$ does not depend on the spatial position, but on the wavevector $\bm{k}$ and the energy $\epsilon$.
We assume that the frequency of the external field $\omega$ is lower than the interband spacing, so that interband transitions are suppressed and negligible.

We calculate the distribution function in the presence of the external field by the semiclassical Boltzmann equation $(\hbar = 1)$ \cite{Luttinger1,Sinitsyn_review}
\begin{equation}
\label{eq:Boltzmann}
\frac{\partial f}{\partial t} + \dot{\bm{k}} \cdot \frac{\partial f}{\partial \bm{k}} = -\mathcal{C}[f].
\end{equation}
The time derivative of the momentum is equal to the force felt by an electron wave packet: $\dot{\bm{k}} = \bm{F}$.
We have an external electric field $\bm{E}(\omega)$, so that the force is $\bm{F} = -e\bm{E}(\omega)$. 
The collision integral $C[f]$ is given by \cite{Luttinger1,Sinitsyn2}
\begin{equation}
\mathcal{C}[f] =
\int_{\bm{k}'}  [w_{\bm{k}'\bm{k}} f(\bm{k}, \epsilon) - w_{\bm{k}\bm{k}'} f(\bm{k}', \epsilon + \delta \epsilon_{\bm{k}'\bm{k}})],
\end{equation}
where $w_{\bm{k}'\bm{k}}$ is the scattering rate from a state with momentum $\bm{k}$ to one with $\bm{k}'$.
In the following, we consider elastic scattering.  Then, the scattering rate becomes 
\begin{equation}
w_{\bm{k}'\bm{k}} = 2\pi |T_{\bm{k}'\bm{k}}| \delta (\epsilon_{\bm{k}'} - \epsilon_{\bm{k}}),
\end{equation}
with the scattering $T$ matrix $T_{\bm{k}'\bm{k}}$.
It is given by $T_{\bm{k}'\bm{k}}=\braket{\bm{k}'|V|\psi_{\bm{k}}}$, where $\ket{\bm{k}}$ is the eigenstate of $H_0$, the Hamiltonian in the clean limit, and $\ket{\psi_{\bm{k}}}$ is the eigenstate of $H_0+V$, including the scattering $V$.  $\ket{\psi_{\bm{k}}}$ can be obtained as the solution to the Lippman--Schwinger equation $\ket{\psi_{\bm{k}}} = \ket{\bm{k}} + (\epsilon_{\bm{k}}-H_0+i\delta)^{-1}V\ket{\psi_{\bm{k}}}$.
The energy shift is obtained as $\delta\epsilon_{\bm{k}'\bm{k}} = -\bm{F} \cdot \delta\bm{r}_{\bm{k}'\bm{k}}$ with the coordinate shift $\delta\bm{r}_{\bm{k}'\bm{k}} \approx (\bm{k}'-\bm{k}) \times \bm{\Omega}(\bm{k})$ for a weak scattering process with a small momentum change \cite{Sinitsyn1}.

\subsection{Scattering rate in noncentrosymmetric media}

In noncentrosymmetric media, the probability of a scattering process $\bm{k}\to\bm{k}'$ and the inverted process $-\bm{k}\to -\bm{k}'$ can be different.  For time-reversal systems, this imbalance is captured by decomposing the scattering rate into two parts:
\begin{equation}
w_{\bm{k}\bm{k}'} = w_{\bm{k}\bm{k}'}^{(S)} + w_{\bm{k}\bm{k}'}^{(A)},
\end{equation}
where the symmetric and antisymmetric parts $w^{(S)}$ and $w^{(A)}$, respectively, is defined by
\begin{equation}
w_{\bm{k}\bm{k}'}^{(S)} = \frac{1}{2} (w_{\bm{k}\bm{k}'} + w_{\bm{k}'\bm{k}}), \quad
w_{\bm{k}\bm{k}'}^{(A)} = \frac{1}{2} (w_{\bm{k}\bm{k}'} - w_{\bm{k}'\bm{k}}).
\end{equation}
When time reversal is preserved, the scattering rate satisfies the reversibility $w_{\bm{k}\bm{k}'} = w_{-\bm{k}',-\bm{k}}$, which leads to the relations
\begin{gather}
w_{\bm{k}'\bm{k}}^{(S)} = w_{-\bm{k},-\bm{k}'}^{(S)} = w_{\bm{k}\bm{k}'}^{(S)}, \\
w_{\bm{k}'\bm{k}}^{(A)} = -w_{-\bm{k},-\bm{k}'}^{(A)} = -w_{\bm{k}\bm{k}'}^{(A)}.
\end{gather}
Those equalities do not hold in general when time reversal is broken since states at $\bm{k}$ and $-\bm{k}$ usually have different energies.
The optical theorem for elastic scattering guarantees the relation
\begin{equation}
\label{eq:optical}
\int_{\bm{k}'} w_{\bm{k}\bm{k}'}^{(A)} = 0.
\end{equation}
When the elastic scattering is due to impurities, $w^{(S)}$ and $w^{(A)}$ are obtained to the lowest order in the impurity scattering potential by
\begin{gather}
\label{eq:scatt_s}
w_{\bm{k}'\bm{k}}^{(S)} = 2\pi \langle |V_{\bm{k}'\bm{k}}|^2 \rangle \delta (\epsilon_{\bm{k}'}-\epsilon_{\bm{k}}), \\
\label{eq:scatt_a}
w_{\bm{k}'\bm{k}}^{(A)} = -(2\pi)^2 \int_{\bm{q}} \operatorname{Im} \langle V_{\bm{k}'\bm{q}} V_{\bm{q}\bm{k}} V_{\bm{k}\bm{k}'} \rangle \delta (\epsilon_{\bm{k}}-\epsilon_{\bm{k}'}) \delta (\epsilon_{\bm{k}'}-\epsilon_{\bm{q}}),
\end{gather}
where $\langle \ \rangle$ denotes the average over the impurity distribution and $V_{\bm{k}'\bm{k}}$ is the matrix element for the single impurity scattering.  We can see that the symmetric part of the scattering rate is obtained at the lowest order of the Born approximation, whereas the antisymmetric part is found at the next leading order.  The detailed balance is broken when the antisymmetric part is finite.

Reference \cite{Belinicher2} reviews the photogalvanic effect in noncentrosymmetric materials, considering the asymmetry of scattering.  The photovoltaic effect is studied with the Boltzmann equation and the antisymmetric component of the scattering rate $w_{\bm{k}'\bm{k}}^{(A)}$, similarly to our analysis.  A difference can be found in the origin of finite $w_{\bm{k}'\bm{k}}$.  In Ref.~\cite{Belinicher2}, the asymmetry is imposed on scattering potentials for impurity scattering, ionization, photoexcitation, and recombination.  We note that the asymmetry of the scattering rate can be finite due to an asymmetric scattering potential and also a wavefunction of a noncentrosymmetric medium.  For the latter case, a scattering potential does not need to be inversion asymmetric to have finite $w_{\bm{k}'\bm{k}}^{(A)}$, but even an isotropic scattering potential can generate $w_{\bm{k}'\bm{k}}^{(A)}$ through a wavefunction.

\subsection{Formal solutions}

The semiclassical Boltzmann equation \eqref{eq:Boltzmann} should be solved self-consistently to obtain the distribution function $f(\bm{k})$.  
A fully self-consistent solution is generally difficult to obtain; here we decompose the distribution function $f$ as
\begin{equation}
f = f_0 + f^\text{scatt} + f^\text{adist}.
\end{equation}
The first term $f_0$ is the distribution function in equilibrium, i.e., the Fermi--Dirac distribution.
The second term $f^\text{scatt}$ describes the scattering contribution without the Berry curvature and the last term $f^\text{adist}$ is the anomalous distribution due to the Berry curvature and the energy shift $\delta\epsilon_{\bm{k}'\bm{k}}$. 
We further expand $f^\text{scatt}$ and $f^\text{adist}$ with respect to the electric field $\bm{E}(\omega)$ and the asymmetric part of the scattering rate $w^{(A)}$:
\begin{gather}
f^\text{scatt}(\bm{k}) = \sum_{n\geq 1} f^\text{scatt}_n(\bm{k}), \quad
f^\text{scatt}_n(\bm{k}) = \sum_{m\geq 0} f_n^{(m)}(\bm{k}), \\
f^\text{adist}(\bm{k}) = \sum_{n\geq 1} f^\text{adist}_n(\bm{k}), \quad
f^\text{adist}_n(\bm{k}) = \sum_{m\geq 0} g_n^{(m)}(\bm{k}).
\end{gather}
The subscript $n$ and the superscript $m$ correspond to the orders of the electric field $E$ and the antisymmetric scattering rate $w^{(A)}$, respectively.
As we have seen in Eqs.~\eqref{eq:scatt_s} and \eqref{eq:scatt_a}, $w^{(A)}$ is smaller than $w^{(S)}$, so that we treat the former as a perturbation.
Then, the Boltzmann equation is decomposed for each $f_n^{(m)}$ or $g_n^{(m)}$ to become
\allowdisplaybreaks[1]
\begin{gather}
\frac{\partial f_1^{(0)}}{\partial t} -e\bm{E}\cdot\frac{\partial f_0}{\partial\bm{k}} = -\int_{\bm{k}'} w_{\bm{k}\bm{k}'}^{(S)} [f_1^{(0)}(\bm{k})-f_1^{(0)}(\bm{k}')], \\
\frac{\partial f_1^{(1)}}{\partial t} = -\int_{\bm{k}'} w_{\bm{k}\bm{k}'}^{(S)} [f_1^{(1)}(\bm{k})-f_1^{(1)}(\bm{k}')] + \int_{\bm{k'}} w_{\bm{k}\bm{k}'}^{(A)} f_1^{(0)}(\bm{k}') , \\
\frac{\partial f_2^{(0)}}{\partial t} -e\bm{E}\cdot\frac{\partial f_1^{(0)}}{\partial\bm{k}} = -\int_{\bm{k}'} w_{\bm{k}\bm{k}'}^{(S)} [f_2^{(0)}(\bm{k})-f_2^{(0)}(\bm{k}')], \\
\frac{\partial f_2^{(1)}}{\partial t} -e\bm{E}\cdot\frac{\partial f_1^{(1)}}{\partial\bm{k}} = -\int_{\bm{k}'} w_{\bm{k}\bm{k}'}^{(S)} [f_2^{(1)}(\bm{k})-f_2^{(1)}(\bm{k}')] + \int_{\bm{k'}} w_{\bm{k}\bm{k}'}^{(A)} f_2^{(0)}(\bm{k}') , \\
\frac{\partial g_1^{(0)}}{\partial t} = -\int_{\bm{k}'} w_{\bm{k}\bm{k}'}^{(S)} [g_1^{(0)}(\bm{k})-g_1^{(0)}(\bm{k}')] + \int_{\bm{k'}} w_{\bm{k}\bm{k}'}^{(S)} f'_0(\bm{k}') (e\bm{E}\cdot\delta\bm{r}_{\bm{k}'\bm{k}}) , \\
\frac{\partial g_1^{(1)}}{\partial t} = -\int_{\bm{k}'} w_{\bm{k}\bm{k}'}^{(S)} [g_1^{(1)}(\bm{k})-g_1^{(1)}(\bm{k}')] + \int_{\bm{k}'} w_{\bm{k}\bm{k}'}^{(A)} g_1^{(0)}(\bm{k}') + \int_{\bm{k'}} w_{\bm{k}\bm{k}'}^{(A)} f'_0(\bm{k}') (e\bm{E}\cdot\delta\bm{r}_{\bm{k}'\bm{k}}) , \\
\begin{aligned}
&\frac{\partial g_2^{(0)}}{\partial t} -e\bm{E}\cdot\frac{\partial g_1^{(0)}}{\partial\bm{k}} \\
=& -\int_{\bm{k}'} w_{\bm{k}\bm{k}'}^{(S)} [g_2^{(0)}(\bm{k})-g_2^{(0)}(\bm{k}')] + \int_{\bm{k'}} w_{\bm{k}\bm{k}'}^{(S)} [f_1^{\prime(0)}(\bm{k}') + g_1^{\prime(1)}(\bm{k}')] (e\bm{E}\cdot\delta\bm{r}_{\bm{k}'\bm{k}}) \\
&+ \frac{1}{2} \int_{\bm{k'}} w_{\bm{k}\bm{k}'}^{(S)} f''_0(\bm{k}') (e\bm{E}\cdot\delta\bm{r}_{\bm{k}'\bm{k}})^2 ,
\end{aligned}\\
\begin{aligned}
&\frac{\partial g_2^{(1)}}{\partial t} -e\bm{E}\cdot\frac{\partial g_1^{(1)}}{\partial\bm{k}} \\
=& -\int_{\bm{k}'} w_{\bm{k}\bm{k}'}^{(S)} [g_2^{(1)}(\bm{k})-g_2^{(1)}(\bm{k}')] + \int_{\bm{k}'} w_{\bm{k}\bm{k}'}^{(S)} [f_1^{\prime(1)}(\bm{k}') + g_1^{\prime(1)}(\bm{k}')] (e\bm{E}\cdot\delta\bm{r}_{\bm{k}'\bm{k}}) \\
&+ \int_{\bm{k'}} w_{\bm{k}\bm{k}'}^{(A)} g_2^{(0)}(\bm{k}') + \int_{\bm{k}'} w_{\bm{k}\bm{k}'}^{(A)} [f_1^{\prime(0)}(\bm{k}') + g_1^{\prime(0)}(\bm{k}')] (e\bm{E}\cdot\delta\bm{r}_{\bm{k}'\bm{k}}) + \frac{1}{2} \int_{\bm{k'}} w_{\bm{k}\bm{k}'}^{(A)} f''_0(\bm{k}') (e\bm{E}\cdot\delta\bm{r}_{\bm{k}'\bm{k}})^2 .
\end{aligned}
\end{gather}
A prime symbol $'$ added to the distribution functions stands for the derivative with respect to the energy: $f'(\bm{k},\epsilon) = \partial_\epsilon f(\bm{k},\epsilon)$.

Each equation has to be solved self-consistently; however it is usually not easy.  Here, we solve a part of the collision integral involving $w^{(S)}$ self-consistently, while the other part with $w^{(A)}$ is treated as a perturbation.  The collision integrals with $w^{(S)}$ define the scattering times $\tau_n^{(m)}$ and $\tau_n^{(m)}$ as follows:
\begin{gather}
\label{eq:scatt_f}
\int_{\bm{k}'} w_{\bm{k}\bm{k}'}^{(S)} [f_n^{(m)}(\bm{k})-f_n^{(m)}(\bm{k}')] = \frac{1}{\tau_{n}^{(m)}} f_n^{(m)}(\bm{k}), \\
\label{eq:scatt_g}
\int_{\bm{k}'} w_{\bm{k}\bm{k}'}^{(S)} [g_n^{(m)}(\bm{k})-g_n^{(m)}(\bm{k}')] = \frac{1}{\tau_{n}^{\prime(m)}} g_n^{(m)}(\bm{k}).
\end{gather}
If $\tau_n^{(m)}$ and $\tau_n^{\prime(m)}$ do not satisfy the self-consistency, one may regard the equations above as the definitions for the relaxation time approximation.
We also note the notation of the scattering times $\tau_n$ and $\tau'_n$ used in the main text; they correspond to $\tau_n \equiv \tau_n^{(0)}$ and $\tau'_n \equiv \tau_n^{(1)}$.

We assume $\tau$ and $\tau'$ are independent of momentum $\bm{k}$ in solving the equations.  Then, we obtain the formal solutions for the distribution functions $f_n^{(m)}$ and $g_n^{(m)}$.  The second-order response contains different frequencies, $0$ and $\pm 2\omega$.  Since now we focus on rectification, i.e., zero-frequency response, we only write down the second-order solutions with $\omega = 0$.  Now, we list below the formal solutions up to the second order in the electric field $(n\leq 2)$ and to the first order in $w^{(A)}$ $(m\leq 1)$:
\begin{gather}
f_1^{(0)}(\bm{k},\omega) = \tau_{1\omega}^{(0)} eE_a \partial_{k_a} f_0(\bm{k}), \\
f_1^{(1)}(\bm{k},\omega) = \tau_{1\omega}^{(1)} \int_{\bm{k}'} w_{\bm{k}\bm{k}'}^{(A)} \tau_{1\omega}^{(0)} eE_a \partial_{k'_a}f_0(\bm{k}'), \\
\label{eq:f20}
f_2^{(0)}(\bm{k},0) = \tau_2^{(0)} eE_a^* \partial_{k_a} f_1^{(0)}(\bm{k},\omega) +\text{c.c.}
= \tau_2^{(0)} eE_a^* \partial_{k_a} \tau_{1\omega}^{(0)} eE_b \partial_{k_b}f_0(\bm{k}) +\text{c.c.} , \\
\label{eq:f21}
\begin{aligned}
&f_2^{(1)}(\bm{k},0) \\
=&\ \tau_2^{(1)} eE_a^* \partial_{k_a} f_1^{(1)}(\bm{k},\omega) + \text{c.c.} + \int_{\bm{k}'} w_{\bm{k}\bm{k'}}^{(A)} f_2^{(0)}(\bm{k}',0) \\
=&\ \tau_2^{(1)} eE_a^* \partial_{k_a} \tau_{1\omega}^{(1)} \int_{\bm{k}'} w_{\bm{k}\bm{k}'}^{(A)} \tau_{1\omega}^{(0)} eE_b \partial_{k'_b} f_0(\bm{k}')
+ \tau_2^{(1)} \int_{\bm{k}'} w_{\bm{k}\bm{k}'}^{(A)} \tau_2^{(0)} eE_a^* \partial_{k'_a} \tau_{1\omega}^{(0)} eE_b \partial_{k'_b} f_0(\bm{k}') + \text{c.c.},
\end{aligned}\\
g_1^{(0)}(\bm{k},\omega) = \tau_{1\omega}^{\prime(0)} \int_{\bm{k}'} w_{\bm{k}\bm{k}'}^{(S)} f'_0(\bm{k}') eE_a \delta r_{a;\bm{k}'\bm{k}}, \\
g_1^{(1)}(\bm{k},\omega) = \tau_{1\omega}^{\prime(1)} \left[ \int_{\bm{k}'} w_{\bm{k}\bm{k}'}^{(A)} \tau_{1\omega}^{\prime(0)} \int_{\bm{k}''} w_{\bm{k}'\bm{k}''}^{(S)} f_0'(\bm{k}'') eE_a \delta r_{a;\bm{k}''\bm{k}'} + \int_{\bm{k}'} w_{\bm{k}\bm{k}'}^{(A)} f_0'(\bm{k}') eE_a \delta r_{a;\bm{k}'\bm{k}} \right], \\
\begin{aligned}
&g_2^{(0)}(\bm{k},0) \\
=&\ \tau_2^{\prime(0)} \bigg[
eE_a^* \partial_{k_a} g_1^{(0)}(\bm{k},\omega)
+\int_{\bm{k}'} w_{\bm{k}\bm{k}'}^{(S)} [ f_1^{\prime(0)}(\bm{k}',\omega) + g_1^{\prime(0)}(\bm{k}',\omega) ] eE_a^* \delta r_{a;\bm{k}'\bm{k}} \\
&+ \frac{1}{2} \int_{\bm{k}'} w_{\bm{k}\bm{k}'}^{(S)} f_0''(\bm{k}') eE_a^* \delta r_{a;\bm{k}'\bm{k}} eE_b \delta r_{b;\bm{k}'\bm{k}}
\bigg] + \text{c.c.} \\
=&\ \tau_2^{\prime(0)} \bigg[
eE_a^* \partial_{k_a} \tau_{1\omega}^{\prime(0)} \int_{\bm{k}'} w_{\bm{k}\bm{k}'}^{(S)} f_0'(\bm{k}') eE_b \delta r_{b;\bm{k}'\bm{k}}
+\int_{\bm{k}'} w_{\bm{k}\bm{k}'}^{(S)} eE_a^* \delta r_{a;\bm{k}'\bm{k}} \partial_{\epsilon_{\bm{k}'}} \tau_{1\omega}^{(0)} eE_b \partial_{k'_b} f_0(\bm{k}') \hspace{5mm}\\
&+\int_{\bm{k}'} w_{\bm{k}\bm{k}'}^{(S)} eE_a^* \delta r_{a;\bm{k}'\bm{k}} \partial_{\epsilon_{\bm{k}'}} \tau_{1\omega}^{\prime(0)} \int_{\bm{k}''} w_{\bm{k}'\bm{k}''}^{(S)} f_0'(\bm{k}'') eE_b \delta r_{b;\bm{k}''\bm{k}'} \\
&+ \frac{1}{2} \int_{\bm{k}'} w_{\bm{k}\bm{k}'}^{(S)} f_0''(\bm{k}') eE_a^* \delta r_{a;\bm{k}'\bm{k}} eE_b \delta r_{b;\bm{k}'\bm{k}}
\bigg] + \text{c.c.},
\end{aligned}\\
\begin{aligned}
& g_2^{(1)}(\bm{k},0) \\
=&\ \tau_2^{\prime(1)} \bigg[
eE_a^* \partial_{k_a} g_1^{(1)}(\bm{k},\omega)
+ \int_{\bm{k}'} w_{\bm{k}\bm{k}'}^{(S)} [ f_1^{\prime(1)}(\bm{k}',\omega)+g_1^{\prime(1)}(\bm{k}',\omega) ] eE_a^* \delta r_{a;\bm{k}'\bm{k}}
+ \int_{\bm{k}'} w_{\bm{k}\bm{k}'}^{(A)} g_2^{(0)}(\bm{k}',0) \\
& + \int_{\bm{k}'} w_{\bm{k}\bm{k}'}^{(A)} [ f_1^{\prime(0)}(\bm{k}',\omega)+g_1^{\prime(0)}(\bm{k}',\omega) ] eE_a^* \delta r_{a;\bm{k}'\bm{k}}
+ \frac{1}{2} \int_{\bm{k}'} w_{\bm{k}\bm{k}'}^{(A)} f_0''(\bm{k}') eE_a^* \delta r_{a;\bm{k}'\bm{k}} eE_b \delta r_{b;\bm{k}'\bm{k}}
\bigg] \\
& + \text{c.c.} \\
=&\ \tau_2^{\prime(1)} \bigg\{
eE_a^* \partial_{k_a} \tau_{1\omega}^{\prime(1)} \left[ \int_{\bm{k}'} w_{\bm{k}\bm{k}'}^{(A)} \tau_{1\omega}^{\prime(0)} \int_{\bm{k}''} w_{\bm{k}'\bm{k}''}^{(S)} f_0'(\bm{k}'') eE_b \delta r_{b;\bm{k}''\bm{k}'} + \int_{\bm{k}'} w_{\bm{k}\bm{k}'}^{(A)} f_0'(\bm{k}') eE_b \delta r_{b;\bm{k}'\bm{k}} \right]
\\
&+ \int_{\bm{k}'} w_{\bm{k}\bm{k}'}^{(S)} eE_a^* \delta r_{a;\bm{k}'\bm{k}}
\partial_{\epsilon_{\bm{k}'}} \bigg[
\tau_{1\omega}^{(1)} \int_{\bm{k}''} w_{\bm{k}'\bm{k}''}^{(A)} \tau_{1\omega}^{(0)} eE_b \partial_{k''_b}f_0(\bm{k}'') \\
&\hspace{20pt} + \tau_{1\omega}^{\prime(1)} \int_{\bm{k}''} w_{\bm{k}'\bm{k}''}^{(A)} \tau_{1\omega}^{\prime(0)} \int_{\bm{k}'''} w_{\bm{k}''\bm{k}'''}^{(S)} f_0'(\bm{k}''') eE_b \delta r_{b;\bm{k}'''\bm{k}''}
+ \tau_{1\omega}^{\prime(1)} \int_{\bm{k}''} w_{\bm{k}'\bm{k}''}^{(A)} f_0'(\bm{k}'') eE_b \delta r_{b;\bm{k}''\bm{k}'}
\bigg] \\
& + \int_{\bm{k}'} w_{\bm{k}\bm{k}'}^{(A)} \tau_2^{\prime(0)} \bigg[
eE_a^* \partial_{k'_a} \tau_{1\omega}^{\prime(0)} \int_{\bm{k}''} w_{\bm{k}'\bm{k}''}^{(S)} f_0'(\bm{k}'') eE_b \delta r_{b;\bm{k}''\bm{k}'} \\
& \hspace{20pt} +\int_{\bm{k}''} w_{\bm{k}'\bm{k}''}^{(S)} eE_a^* \delta r_{a;\bm{k}''\bm{k}'} \partial_{\epsilon_{\bm{k}''}} \tau_{1\omega}^{(0)} eE_b \partial_{k''_b} f_0(\bm{k}'') \\
& \hspace{20pt} +\int_{\bm{k}''} w_{\bm{k}'\bm{k}''}^{(S)} eE_a^* \delta r_{a;\bm{k}''\bm{k}'} \partial_{\epsilon_{\bm{k}''}} \tau_{1\omega}^{\prime(0)} \int_{\bm{k}'''} w_{\bm{k}''\bm{k}'''}^{(S)} f_0'(\bm{k}''') eE_b \delta r_{b;\bm{k}'''\bm{k}''} \\
& \hspace{20pt} + \frac{1}{2} \int_{\bm{k}''} w_{\bm{k}'\bm{k}''}^{(S)} f_0''(\bm{k}'') eE_a^* \delta r_{a;\bm{k}''\bm{k}'} eE_b \delta r_{b;\bm{k}''\bm{k}'}
\bigg] \\
& + \int_{\bm{k}'} w_{\bm{k}\bm{k}'}^{(A)} eE_a^* \delta r_{a;\bm{k}'\bm{k}} \left[ \tau_{1\omega}^{(0)} eE_b \partial_{k_a} f'_0(\bm{k}) + \partial_{\epsilon_{\bm{k}'}} \tau_{1\omega}^{\prime(0)} \int_{\bm{k}''} w_{\bm{k}'\bm{k}''}^{(S)} f'_0(\bm{k}'') eE_b \delta r_{b;\bm{k}''\bm{k}'} \right] \\
& + \frac{1}{2} \int_{\bm{k}'} w_{\bm{k}\bm{k}'}^{(A)} f_0''(\bm{k}') eE_a^* \delta r_{b;\bm{k}'\bm{k}} eE_b \delta r_{b;\bm{k}'\bm{k}}
\bigg\} + \text{c.c.}
\end{aligned}
\end{gather}
$\tau_{n\omega}^{(m)}$ and $\tau_{n\omega}^{\prime(m)}$ are the shorthand notations for
\begin{equation}
\tau_{n\omega}^{(m)} = \frac{\tau_n^{(m)}}{1-i\omega\tau_n^{(m)}}, \quad
\tau_{n\omega}^{\prime(m)} = \frac{\tau_n^{\prime(m)}}{1-i\omega\tau_n^{\prime(m)}}.
\end{equation}
Those formal solutions find the scattering time dependence of the distribution functions for low frequencies $(\omega\tau\ll1)$:
\begin{gather}
f_n^\text{scatt} \propto \tau^n E^n, \nonumber \\
f_n^\text{adist} \propto (\tau^0 E^n,\tau^1 E^n,\ldots, \tau^{n-1}E^n).
\end{gather}
Higher-order corrections for the anomalous distribution $(n\geq 2)$ has a nested structure of $f_{n-1}^\text{scatt}$ and $f_{n-1}^\text{adist}$.
In the weak disorder limit of $\tau\to\infty$, we obtain
\begin{equation}
f_n^\text{adist} \propto \tau^{n-1}E^n.
\end{equation}
We can define another scattering time $\tilde{\tau}$ from the antisymmetric part $w^{(A)}$.  It is roughly speaking given by
\begin{equation}
\label{eq:time_asym}
\int_{\bm{k}'} w_{\bm{k}\bm{k}'}^{(A)} f(\bm{k}') \sim \frac{1}{\tilde{\tau}} f(\bm{k}).
\end{equation}
We note that the definition of $\tilde{\tau}$ is accompanied with the distribution function because the optical theorem for elastic scattering concludes Eq.~\eqref{eq:optical}.

\section{Current response}

From the distribution function, the electric current response is obtained by
\begin{equation}
\bm{j} = -e \int_{\bm{k}} \bm{v}(\bm{k}) f(\bm{k}).
\end{equation}
$\bm{v}(\bm{k})$ is the electron's group velocity, given by
\begin{align}
\bm{v} &= \frac{\partial\epsilon_{\bm{k}}}{\partial\bm{k}} - \dot{\bm{k}} \times \bm{\Omega} + \int_{\bm{k}'} w_{\bm{k}'\bm{k}} \delta\bm{r}_{\bm{k}'\bm{k}} \nonumber\\
&\equiv \bm{v}_0 + \bm{v}_\text{av} + \bm{v}_\text{sj}.
\end{align}
The group velocity from the energy band dispersion $\bm{v}_0$ is independent of the Berry curvature, while the Berry curvature induces the anomalous velocity $\bm{v}_\text{av}$ \cite{Karplus,Luttinger3,Adams} and the side-jump velocity $\bm{v}_\text{sj}$ \cite{Berger}: 
\begin{gather}
\bm{v}_\text{av}=-\dot{\bm{k}}\times\bm{\Omega}=e\bm{E}\times\bm{\Omega}, \\
\bm{v}_\text{sj}=\int_{\bm{k}'}w_{\bm{k}'\bm{k}}\delta\bm{r}_{\bm{k}'\bm{k}}. 
\end{gather}
In terms of the electric field $E$, $\bm{v}_\text{av}$ is linear in $E$ whereas $\bm{v}_0$ and $\bm{v}_\text{sj}$ do not depend on $E$. Lastly, only the side-jump velocity depends on the scattering time: $\bm{v}_\text{sj}\propto \tau^{-1}$.
Since there is no current in equilibrium, the electric current is described by \cite{AHE,Sinitsyn_review}
\begin{equation}
\label{eq:current}
\bm{j} = -e\int_{\bm{k}} (\bm{v}_0 + \bm{v}_\text{av} + \bm{v}_\text{sj}) (f^\text{scatt} + f^\text{adist}),
\end{equation}
which allows us to decompose the current into contributions of different origins.  We note that the contribution from the anomalous distribution $f^\text{adist}$ could be recognized as a part of the side-jump effect since it is originated both from the Berry curvature and scattering.

Before calculating the second-order conductivity, we note the problem about the energy dissipation or Joule heating because of $\bm{j}\cdot\bm{E}\neq 0$; see also e.g., Ref.~\cite{Luttinger1} for discussions.  This concludes that a stationary state solution cannot be obtained for a closed system with energy conservation.  To circumvent this issue, we neglect parts of the distribution function which are isotropically coupled to the electric field.  To be more explicit, the distribution function at the second order in the electric field involves a term proportional to $|\bm{E}|^2 f''_0$.
Technically, such an excess term arises when we substitute a distribution function $f_n$ $(n\geq 1)$ into the collision integral $\mathcal{C}[f]$, which involves only elastic scattering.  Since this term is isotropic, it cannot be relaxed by elastic scattering, which conserves energy, so that a stationary state solution does not exist.  Inelastic scattering resolves the problem as it does not conserve energy and excess energy is dissipated as heat.  Note that an inelastic scattering time is typically much longer than an elastic scattering time.
In calculating the second-order conductivity, we simply subtract and neglect such isotropic terms although they potentially change the temperature for a closed system.
Importantly, they do not contribute to current from the band velocity since $\int_{\bm{k}} \bm{v}_0 f''_0=\int_{\bm{k}} \nabla_{\bm{k}} f'_0=0$.

We can see the six possible combinations in Eq.~\eqref{eq:current}.
For the second-order current response, the scattering time dependence of each contribution for low frequencies $\omega\tau\ll1$ is
\begin{gather}
\bm{v}_0 f^\text{scatt}_2 \propto \tau^2, \quad
\bm{v}_\text{av} f^\text{scatt}_1 \propto \tau^1, \quad
\bm{v}_\text{sj} f^\text{scatt}_2 \propto \tau^1, \\
\bm{v}_0 f^\text{adist}_2 \propto (\tau^1, \tau^0), \quad
\bm{v}_\text{av} f^\text{adist}_1 \propto \tau^0, \quad
\bm{v}_\text{sj} f^\text{adist}_2 \propto (\tau^0, \tau^{-1}).
\end{gather}
In the weak disorder limit of $\tau\to\infty$, the first term, the skew scattering contribution, $\bm{v}_0 f^\text{scatt}$ predominates, which we focus on in this work.  This is equivalent to neglect all Berry curvature related effects.
The second-order current response arising from this term is
\begin{align}
\bm{j}_2 = -e\int_{\bm{k}} \bm{v}_0(\bm{k}) f^\text{scatt}_2(\bm{k})
= -e\int_{\bm{k}} \bm{v}_0(\bm{k}) [f_2^{(0)}(\bm{k}) + f_2^{(1)}(\bm{k})] .
\end{align}
We define the second-order conductivity by the relation
\begin{equation}
j_{2,a} = \chi_{abc} E_b^* E_c,
\end{equation}
and we obtain from Eqs.~\eqref{eq:f20} and \eqref{eq:f21}
\begin{gather}
\chi_{abc} = \chi_{abc}^{(0)} + \chi_{abc}^{(1)}, \\
\label{eq:chi0}
\chi_{abc}^{(0)} = -e^3 \tau_2^{(0)}\tau_{1\omega}^{(0)} \int_{\bm{k}} v_{0,a} \partial_{k_b} \partial_{k_c} f_0(\bm{k}) + (b \leftrightarrow c)^* ,  \\
\begin{aligned}
\label{eq:chi1}
\chi_{abc}^{(1)} = &-e^3 \bigg[ \tau_2^{(1)}\tau_{1\omega}^{(1)}\tau_{1\omega}^{(0)} \int_{\bm{k}} v_{0,a} \partial_{k_b} \int_{\bm{k}'} w_{\bm{k}\bm{k}'}^{(A)} \partial_{k'_c} f_0(\bm{k}') \\
&\hspace{20pt} + \tau_2^{(1)}\tau_2^{(0)}\tau_{1\omega}^{(0)}
\int_{\bm{k}} v_{0,a} \int_{\bm{k}'} w_{\bm{k}\bm{k}'}^{(A)} \partial_{k'_b} \partial_{k'_c} f_0(\bm{k}')
\bigg] \\
&+ (b \leftrightarrow c)^*.
\end{aligned}
\end{gather}
$\chi_{abc}^{(0)}$ vanishes identically in time-reversal-invariant systems because of $\epsilon_{\bm{k}} = \epsilon_{-\bm{k}}$; it follows from $\bm{v}_0(\bm{k}) = -\bm{v}_0(-\bm{k})$, and we see that the integrand is odd in $\bm{k}$ for $\chi_{abc}^{(0)}$.  Therefore, the second-order contribution from the skew-scattering contribution arises from $\chi_{abc}^{(1)}$, i.e., $\chi_{abc} = \chi_{abc}^{(1)}$.

To calculate the second-order response, it is useful to express Eqs.~\eqref{eq:chi0} and \eqref{eq:chi1} in the following forms:
\begin{gather}
\begin{aligned}
\chi_{abc}^{(0)} &= -e^3 \tau_2^{(0)} \tau_{1\omega}^{(0)} \int_{\bm{k}} [-f'_0(\bm{k})] v_{0,a} (\partial_{k_b}\partial_{k_c}\epsilon_{\bm{k}}) + (b \leftrightarrow c)^* \\
&= -2e^3 \operatorname{Re} \left[ \tau_2^{(0)}\tau_{1\omega}^{(0)} \right] \int_{\bm{k}} [-f'_0(\bm{k})] v_{0,a} (\partial_{k_b}\partial_{k_c}\epsilon_{\bm{k}})
,
\end{aligned}\\
\begin{aligned}
\chi_{abc}^{(1)} =& -e^3 \Bigg[
\tau_2^{(1)}\tau_{1\omega}^{(1)}\tau_{1\omega}^{(0)} \int_{\bm{k}} [-f'_0(\bm{k})] (\partial_{k_a}\partial_{k_b}\epsilon_{\bm{k}}) \int_{\bm{k}'} w_{\bm{k}\bm{k}'}^{(A)} v_{0,c} \\
&\hspace{20pt} -\tau_2^{(1)}\tau_2^{(0)}\tau_{1\omega}^{(0)} \int_{\bm{k}} [-f'_0(\bm{k})] v_{0,a} \int_{\bm{k}'} w_{\bm{k}\bm{k}'}^{(A)} (\partial_{k_b}\partial_{k_c}\epsilon_{\bm{k}}) \\
&\hspace{20pt} +\tau_2^{(1)}\tau_2^{(0)}\tau_{1\omega}^{(0)} \int_{\bm{k}} [-f'_0(\bm{k})] \partial_{k_a} \int_{\bm{k}'} w_{\bm{k}\bm{k}'}^{(A)} v_{0,b} v_{0,c}
\Bigg]\\
&+ (b \leftrightarrow c)^*.
\end{aligned}
\end{gather}
We also show the expression for $\chi_{abc}^{(0)}$ for reference although it vanishes.
These expressions show that results at finite temperature can be obtained by considering the thermal broadening of the Fermi--Dirac distribution, where the Fermi energy $\epsilon_F$ is replaced by the chemical potential $\mu_\text{chem}$ with the carrier density fixed.  More explicitly, from the following relation
\begin{equation}
-f'_0(\bm{k}) = \int d\epsilon [-f'_0(\epsilon;\mu_\text{chem})] \delta (\epsilon_{\bm{k}}-\epsilon_F),
\end{equation}
the second-order conductivity at finite temperature is obtained by
\begin{equation}
\chi_{abc}(\mu_\text{chem},T) = \int d\epsilon [-f'_0(\epsilon;\mu_\text{chem})] \chi_{abc}(\epsilon,T=0).
\end{equation}

\subsection{Linear response}

We also calculate the linear response to obtain responsivities (defined below).  From the Boltzmann equation, the linear current response is given by
\begin{equation}
j_{1,a}(\omega) = \sigma_{ab}E_b(\omega) = -e\int_{\bm{k}} v_{0,a} f_1^\text{scatt},
\end{equation}
leading to the linear conductivity
\begin{gather}
\sigma_{ab} = \sigma_{ab}^{(0)} + \sigma_{ab}^{(1)}, \\
\label{eq:linear_0}
\sigma_{ab}^{(0)} = -e^2 \tau_{1\omega}^{(0)} \int_{\bm{k}} v_{0,a} \partial_{k_b} f_0(\bm{k})
= e^2 \tau_{1\omega}^{(0)}\int_{\bm{k}} v_{0,a}v_{0,b} [-f'_0(\bm{k})], \\
\sigma_{ab}^{(1)} = -e^2 \tau_{1\omega}^{(1)}\tau_{1\omega}^{(0)} \int_{\bm{k}} v_{0,a} \int_{\bm{k}'} w_{\bm{k}\bm{k}'}^{(A)} \partial_{k_b'} f_0(\bm{k}').
\end{gather}

\section{Graphene-based models with trigonal lattice structures}

Now we evaluate the current response for explicit models.  We consider two graphene-based models, utilizing monolayer and bilayer graphene.  The crystalline lattices belong to a crystallographic point group $D_{3h}$ and $C_{3v}$, respectively.  We note that inversion is equivalent to the in-plane two-fold rotation in 2D models, which is absent in the present models.  Around $K$ and $K'$ points, $\bm{K}=\left(\frac{4\pi}{3\sqrt{3}},0\right)$ and $\bm{K}'=\left(-\frac{4\pi}{3\sqrt{3}},0\right)$, the $k\cdot p$ Hamiltonian to second order is
\begin{align}
H_s(\bm{k}) &=
\begin{pmatrix}
\Delta & svk_{-s} -\lambda k_s^2 \\
svk_s - \lambda k_{-s}^2 & -\Delta
\end{pmatrix} \nonumber\\
&=
\left[ svk_x -\lambda (k_x^2-k_y^2) \right] \sigma_x + \left( vk_y + 2s\lambda k_x k_y \right) \sigma_y + \Delta\sigma_z,
\label{eq:hamiltonian_d3h}
\end{align}
where $k_x$ and $k_y$ are measured from $K$ or $K'$ point, labeled by $s=\pm1$, and $k_s$ is defined by $k_s=k_x+isk_y$.  The energy dispersion of the positive branch is given by 
\begin{equation}
\epsilon_{\bm{k}} = \sqrt{v^2 k^2 + \lambda^2 k^4 + \Delta^2 -2sv\lambda k^3 \cos 3\theta}, 
\end{equation}
and the corresponding normalized wavefunction is 
\begin{equation}
\ket{\bm{k},s} = \frac{1}{\sqrt{2\epsilon_{\bm{k}}(\epsilon_{\bm{k}}+\Delta)}}
\begin{pmatrix}
\epsilon_{\bm{k}}+\Delta \\
\left[svk_x -\lambda (k_x^2-k_y^2)\right]+ i\left( vk_y + 2s\lambda k_x k_y \right)
\end{pmatrix}.
\end{equation}
The Berry curvature is also calculated as 
\begin{equation}
\Omega_z = -\frac{s\Delta(v^2-4\lambda^2k^4)}{2\epsilon_{\bm{k}}^3}. 
\end{equation}

\begin{figure}
\centering
\includegraphics[width=0.8\hsize]{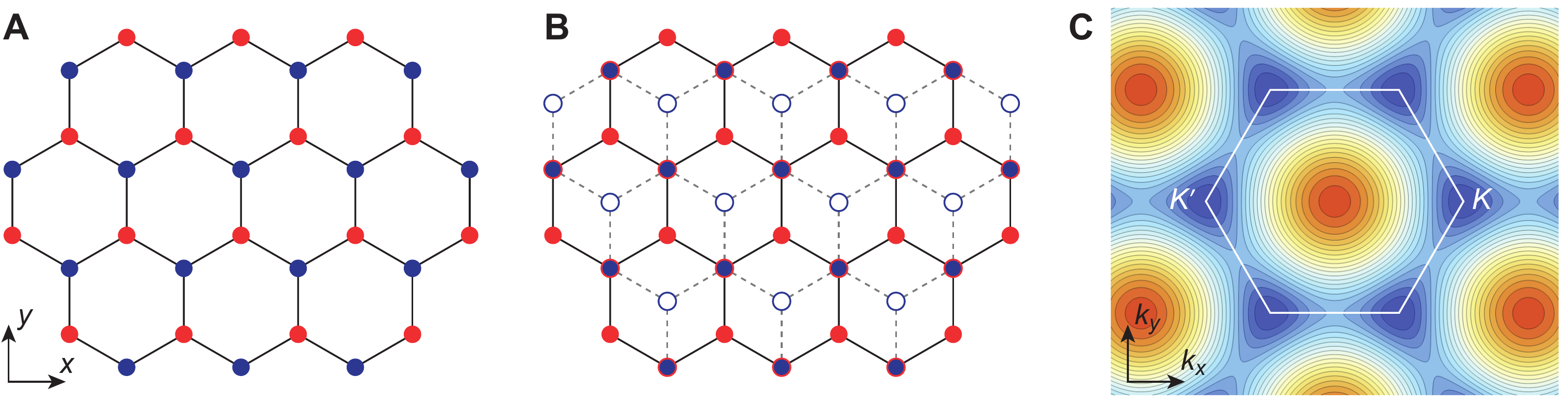}
\caption{
\textbf{Monolayer and bilayer graphene models.}
(\textbf{A}) Lattice structures of monolayer graphene and (\textbf{B}) bilayer graphene along with the crystal axes.  The red and blue circles depict distinct sublattice sites with different potential energies, and the filled and empty circles correspond to atoms on different layers.  Both lattices have reflection planes parallel to the $y$ axis and its symmetry partners under the three-fold rotation.
(\textbf{C}) Brillouin zone and energy contours for the monolayer model.  With small doping, there are trigonally-warped Fermi surfaces around $K$ and $K'$ points, facing the opposite directions.  The bilayer model exhibits the similar Fermi surfaces as the monolayer ones.
}
\label{fig:graphene}
\end{figure}

The Hamiltonian Eq.~\eqref{eq:hamiltonian_d3h} is available to both monolayer and bilayer models (Fig.~\ref{fig:graphene}).
The Pauli matrix $\sigma_a$ describes the two sublattices for monolayer graphene and two layers for AB (Bernal) stacked bilayer graphene.  Another difference is found in which of $v$ and $\lambda$ is dominant, which is explained later.  $\Delta$ describes the potential energy difference at the two sublattice sites or layers, which opens a gap to the energy spectrum.
In the following, we consider the two cases separately.

\subsection{Monolayer graphene}

For monolayer graphene, the terms with $v$ dominate those with $\lambda$ $(vk_F \gg \lambda k_F^2)$, where $\lambda$ characterizes the trigonal warping of the Fermi surface.
$s=+1$ corresponds to $K$ valley and $s=-1$ to $K'$ valley.
The Dirac mass can be induced for example by placing a graphene sheet on a hexagonal boron nitride (hBN) substrate.
For a tight-binding model on a honeycomb lattice with the nearest-neighbor hopping amplitude $t$ and the distance between sites $a$, the velocity $v$ and the trigonal warping $\lambda$ are given by $v=3ta/2$ and $\lambda=3ta^2/8$, respectively \cite{graphene_review1,graphene_review2}.

We assume that the trigonal warping is small and obtain the energy dispersion $\epsilon_{\bm{k}}$ to first order in $\lambda$ as
\begin{equation}
\epsilon_{\bm{k}} = \sqrt{v^2k^2 + \Delta^2} -sv\lambda \frac{k^3}{\sqrt{v^2k^2+\Delta^2}} \cos 3\theta_{\bm{k}} + O(\lambda^2),
\end{equation}
where the Fermi wavevector $k_F$ is given by
\begin{equation}
k_F = \frac{\sqrt{\epsilon_F^2-\Delta^2}}{v}.
\end{equation}
The density of states (DOS) at the Fermi energy is
\begin{equation}
D_0 = \frac{\epsilon_F}{2\pi v^2}, 
\end{equation}
and $\theta_{\bm{k}}$ denotes the polar angle of $\bm{k}$. 
The Fermi surface can be parameterized by $\theta_{\bm{k}}$ as 
\begin{equation}
k(\theta_{\bm{k}}) = k_F + s\lambda \frac{k_F^2}{v} \cos 3\theta_{\bm{k}}.
\end{equation}
To evaluate the scattering rate, we need the wavefunction on the Fermi surface. The normalized wavefunction on the Fermi surface is given by
\begin{equation}
\ket{\bm{k},s} = \left( \sqrt{\frac{\epsilon_F+\Delta}{2\epsilon_F}}, \frac{(s\cos\theta_{\bm{k}}+i\sin\theta_{\bm{k}})(vk_F+i\lambda k_F^2\sin 3\theta_{\bm{k}})}{\sqrt{2\epsilon_F(\epsilon_F+\Delta)}} \right)^T .
\end{equation}
Also, it is useful to see the Berry curvature on the Fermi surface, which is given by 
\begin{equation}
\Omega_z = -\frac{sv^2\Delta}{2\epsilon_F^3} + O(\lambda^2). 
\end{equation}

The second-order conductivity is considered in the presence of an elastic impurity scattering.  Here, we assume short-ranged random scalar impurities
\begin{equation}
\label{eq:scattering}
V(\bm{r}) = V_i \sum_j \delta(\bm{r}-\bm{r}_j),
\end{equation}
but restrict scattering within each valley.
$V_i$ is the impurity potential strength, $n_i$ is the impurity density, and $\bm{r}_j$ denotes an impurity position, where the summation is taken over all impurity sites.
Then, we obtain the scattering rates for the valley $s$ to first order in $\lambda$ as
\begin{gather}
\begin{aligned}
w_{s,\bm{k}\bm{k}'}^{(S)} = \frac{\pi n_i V_i^2}{2\epsilon_F^2 (\epsilon_F+\Delta)^2} \Big[ & |(\epsilon_F+\Delta)^2 + v^2k_F^2 e^{i(\theta-\theta')}|^2 \\
& -2s\lambda vk_F^3 (\epsilon_F+\Delta)^2 \sin(\theta-\theta') (\sin 3\theta-\sin 3\theta') \Big]
\delta(\epsilon_{\bm{k}}-\epsilon_{\bm{k}'}),
\end{aligned} \\
\begin{aligned}
\label{eq:scatter_asym_1}
w_{s,\bm{k}\bm{k}'}^{(A)} = & -\frac{\pi n_i V_i^3 \Delta k_F^2}{2\epsilon_F^2} \Big\{ s\sin (\theta-\theta') \\
&\hspace{30pt} +\lambda\frac{k_F}{v} [ \sin(\theta-4\theta') + \sin(4\theta-\theta') - \sin(\theta+2\theta') + \sin(2\theta+\theta') ]
\Big\}
\delta(\epsilon_{\bm{k}}-\epsilon_{\bm{k}'}),
\end{aligned}
\end{gather}
with $\theta=\theta_{\bm{k}}$ and $\theta'=\theta_{\bm{k}'}$ parametrizing the Fermi surface.

In order to obtain the distribution function, we need to determine the scattering times that satisfy the relation Eq.~\eqref{eq:scatt_f}.  For the present Hamiltonian, the scattering times are obtained as
\begin{gather}
\tau_1^{(0)} = \tau_1^{(1)} = \tau_2^{(1)}
= \left( n_i V_i^2 \frac{\epsilon_F^2 + 3\Delta^2}{4v^2 \epsilon_F} \right)^{-1}, \\
\tau_1^{(1)}
= \left( n_i V_i^2 \frac{\epsilon_F^2 + \Delta^2}{2v^2 \epsilon_F} \right)^{-1}.
\end{gather}
These solutions are obtained to $O(\lambda^0)$, which are enough to evaluate conductivities to order $O(\lambda^1)$ (see the next paragraph).  Also, there is no directional dependence in the scattering times at order $\lambda^0$.  For clarity, we define
\begin{equation}
\tau = \left( n_i V_i^2 \frac{\epsilon_F^2 + 3\Delta^2}{4v^2 \epsilon_F} \right)^{-1},
\end{equation}
and the dimensionless quantity
\begin{equation}
\gamma = \frac{\epsilon_F^2 + 3\Delta^2}{2(\epsilon_F^2 + \Delta^2)} (\leq1),
\end{equation}
which are defined to satisfy $\tau_1^{(0)} = \tau_1^{(1)} = \tau_2^{(1)} = \tau$ and $\tau_1^{(1)} = \gamma\tau$.

We calculate the second-order contribution from each valley separately, namely $\chi_{abc} = \chi_{abc}^{(0)} + \chi_{abc}^{(1)}$ with
\begin{equation}
\chi_{abc}^{(0)} = \sum_s \chi_{s,abc}^{(0)}, \quad
\chi_{abc}^{(1)} = \sum_s \chi_{s,abc}^{(1)}.
\end{equation}
The contributions from each valley $s$ are obtained from Eqs.~\eqref{eq:chi0} and \eqref{eq:chi1} at zero temperature as
\begin{gather}
\chi_{s,xxx}^{(0)} = -\chi_{s,xyy}^{(0)} = -\chi_{s,yxy}^{(0)} = -\chi_{s,yyx}^{(0)}
= s e^3 \tau^2 v \frac{3\lambda k_F^2(\epsilon_F^2+\Delta^2) \gamma}{\pi\epsilon_F^3} \operatorname{Re} \left[\frac{1}{1-i\omega\tau}\right], \\ 
\label{eq:result2}
\chi_{s,xxy}^{(1)} = \chi_{s,xyx}^{(1)} = \chi_{s,yxx}^{(1)} = -\chi_{s,yyy}^{(1)}
= 4e^3 \tau^2 v \frac{\tau}{\tilde{\tau}} \operatorname{Re}\left[ \frac{vk_F(2\epsilon_F^2+\Delta^2)}{(1-i\omega\tau)^2 \pi\epsilon_F^3} -\gamma \frac{vk_F(\epsilon_F^2+2\Delta^2)}{(1-i\omega\tau)\pi\epsilon_F^3} \right],
\end{gather}
where the factor of two from spin is multiplied.
We note that $D_{3h}$ symmetry concludes that the imaginary part of the second-order DC conductivity vanishes \cite{Belinicher2} and that the system does not have second-order response to a circularly-polarized field.
We define another scattering time $\tilde{\tau}$, related to $w^{(A)}$ [see Eq.~\eqref{eq:time_asym}], as
\begin{equation}
\tilde{\tau} = \left( \Delta\lambda \frac{n_i V_i^3 k_F^3}{8v^3\epsilon_F} \right)^{-1}.
\end{equation}
This quantity measures the amounts of inversion breaking $(m)$ and the Fermi surface warping $(\lambda)$.
The ratio $\tau/\tilde{\tau}$ becomes
\begin{equation}
\frac{\tau}{\tilde{\tau}} = \Delta\lambda \frac{V_i k_F^3}{2v(\epsilon_F^2+3\Delta^2)}
= \frac{1}{2} \frac{\Delta \cdot V_i k_F^2}{\epsilon_F^2+3\Delta^2} \frac{\lambda k_F^2}{vk_F}.
\end{equation}
Roughly speaking, the ratio is determined by the product of the following dimensionless quantities: inversion breaking $\Delta/\epsilon_F$, the Fermi surface warping $\lambda k_F^2/(vk_F)$, and the impurity strength $V_ik_F^2/\epsilon_F$.
We can see from Eq.~\eqref{eq:scatter_asym_1} that the antisymmetric scattering becomes finite when inversion is broken by $\Delta \neq 0$ but it does not require finite Fermi surface warping $\lambda$. However, the second-order response vanishes when the Fermi surface warping is absent.
The ratio $\tau/\tilde{\tau}$ measures the amount of the contribution to the second-order response from skew scattering, but not the magnitude of skew scattering itself.

$\chi^{(0)}$ is originated from the absence of time-reversal or inversion symmetry within each valley, resulting from the Fermi surface anisotropy.  However, time reversal is preserved with the two valleys, and thus contributions from different valleys cancel.  
This effect of $\chi^{(0)}$ can be seen in Fig.~2(c), where the two valleys generates the opposite static velocity $\bm{V}$. 
In contrast, $\chi^{(1)}$ adds up with the two valleys, since this contribution captures the opposite chirality of the wavefunctions around each valley, which is not visible from the Fermi surface shape.
Finally, we obtain the nonvanishing elements of the second-order conductivity
\begin{gather}
\label{eq:result-graphene}
\chi_{xxy} = \chi_{xyx} = \chi_{yxx} = -\chi_{yyy} = -4e^3 v_F \frac{\tau^3}{\tilde{\tau}} \zeta(\omega) \equiv -\chi(\omega),
\end{gather}
with the average Fermi velocity on the Fermi surface
\begin{equation}
v_F = \frac{v^2k_F}{\epsilon_F},
\end{equation}
and the dimensionless function
\begin{gather}
\zeta(\omega) = \frac{2}{\pi} \operatorname{Re} \left[ \gamma \frac{\epsilon_F^2+2\Delta^2}{(1-i\omega\tau)\epsilon_F^2} - \frac{2\epsilon_F^2+\Delta^2}{(1-i\omega\tau)^2 \epsilon_F^2} \right].
\end{gather}
Here the factor of four in Eq.~\eqref{eq:result-graphene} corresponds to spin and valley degrees of freedom.
The finite elements of $\chi_{abc}$ are consistent with $D_{3h}$ symmetry of the lattice.

\subsubsection{Linear conductivity and responsivity}

To estimate the magnitude of the second-order response and to calculate responsivities, we also calculate the linear conductivity, particularly $\sigma_{s,ab}^{(0)}$.  From Eq.~\eqref{eq:linear_0}, we obtain
\begin{equation}
\sigma_{s,ab}^{(0)} = 2e^2 \tau_{1\omega}^{(0)} \frac{v^2k_F^2}{4\pi\epsilon_F} \delta_{ab},
\end{equation}
for each valley with the spin degrees of freedom included.
The linear conductivity is diagonal, and hence we write
\begin{equation}
\sigma(\omega) = \sum_s \operatorname{Re} \sigma_{s,aa}^{(0)}
= e^2 \tau \frac{v^2k_F^2}{\pi\epsilon_F} \frac{1}{1+(\omega\tau)^2}.
\end{equation}
Since the linear conductivity can be written as $\sigma = ne\mu$ with the electron density $n$ and the mobility $\mu$, in low frequencies we find the relation
\begin{equation}
\mu = e\tau \frac{v^2}{\epsilon_F}.
\end{equation}
We note that the electron density is obtained by $n= k_F^2/\pi$.

The short-circuit current responsivity $\mathfrak{R}_I$ is defined as the ratio of the generated DC current $j_2$ to the power of the incident electric field absorbed by the sample.  When a sample has a dimension of $L^2$ and the incident field is uniform on the sample, the current responsivity is given by
\begin{equation}
\mathfrak{R}_I \equiv \frac{j_2 L}{j_1 E L^2} = \frac{1}{L} \frac{\chi(\omega)}{\sigma(\omega)}.
\end{equation}
We note that the dimension $L$ corresponds to either width or length of a sample, depending on whether the second-order response of interest is longitudinal or transverse to the incident electric field.
We also define the voltage responsivity $\mathfrak{R}_V$ as the ratio of the voltage generated by the second-order response to the incident power:
\begin{equation}
\mathfrak{R}_V = \frac{j_2L/\sigma(0)}{j_1 E L^2} = \frac{1}{L} \frac{\chi(\omega)}{\sigma(\omega) \sigma(0)}.
\end{equation}
Both current responsivity $\mathfrak{R}_I$ and voltage responsivity $\mathfrak{R}_V$ depend on the sample dimension $L$ and the ratio $\tau/\tilde{\tau}$.  To remove those factors, we define the reduced current responsivity $\eta_I$ and the reduced voltage responsivity $\eta_V$ as follows:
\begin{gather}
\mathfrak{R}_V = \eta_I \frac{1}{L} \frac{\tau}{\tilde{\tau}}, \\
\mathfrak{R}_I = \eta_V \frac{1}{L} \frac{\tau}{\tilde{\tau}}.
\end{gather}
For the present model, $\eta_I$ and $\eta_V$ are obtained as
\begin{gather}
\eta_I = \frac{4\pi e\tau}{k_F} [ 1+(\omega\tau)^2 ] \zeta(\omega), \\
\eta_V = \frac{4\pi}{env_F} \left[ 1+(\omega\tau)^2 \right] \zeta(\omega).
\end{gather}

\subsection{Bilayer graphene}

Bernal stacked bilayer graphene has $C_{3v}$ symmetry when the two layer have different potential energies.  Because of the symmetry, we can use the same Hamiltonian Eq.~\eqref{eq:hamiltonian_d3h}; however, $v$ is treated as a perturbation instead of $\lambda$ for the bilayer case $(\lambda k_F^2 \gg vk_F)$, as we can see from a tight-binging Hamiltonian \cite{graphene_review1,graphene_review2}.
Treating $v$ as a perturbation, we obtain the energy dispersion $\epsilon_{\bm{k}}$ to first order in $v$ as
\begin{equation}
\epsilon_{\bm{k}} = \sqrt{ \lambda^2 k^4 + \Delta^2} - \frac{2s v\lambda k^3}{\sqrt{ \lambda^2 k^4 + \Delta^2 \cos 3\theta_{\bm{k}}}}.
\end{equation}
The DOS at the Fermi energy $D_0$ is given by
\begin{equation}
D_0 = \frac{\epsilon_F}{4\pi k_F^2}.
\end{equation}
The normalized wavefunction on the Fermi surface to order $v$ is
\begin{equation}
\ket{\bm{k},s} = \left( \sqrt{\frac{\epsilon_F+\Delta}{2\epsilon_F}}, -\frac{(\cos2\theta_{\bm{k}}-is\sin2\theta_{\bm{k}})(\lambda k_F^2-ivk_F \sin 3\theta_{\bm{k}})}{\sqrt{2\epsilon_F(\epsilon_F+\Delta)}} \right)^T,
\end{equation}
where $k_F$ is defined by
\begin{equation}
k_F = \left( \frac{\epsilon_F^2-\Delta^2}{\lambda^2} \right)^{1/4}.
\end{equation}
The Berry curvature on the Fermi surface is 
\begin{equation}
\Omega_z = \frac{2s\Delta\lambda^2 k_F^2}{\epsilon_F^3} + \frac{2\Delta v\lambda k_F \cos 3\theta}{\epsilon_F^3} +O(v^2). 
\end{equation}
Using the wavefunction, we obtain the symmetric and antisymmetric scattering rates $w^{(S)}$ and $w^{(A)}$, respectively:
\begin{gather}
\begin{aligned}
w_{s,\bm{k}\bm{k}'}^{(S)} = \ &\frac{\pi n_i V_i^2}{2\epsilon_F^2 (\epsilon_F+ \Delta)^2} \bigg[ \big|(\epsilon_F+\Delta)^2 \\
& + \lambda^2 k_F^4 e^{2i(\theta-\theta')} \big|^2 - 2sv\lambda (\epsilon_F+\Delta)^2 k_F^3 \sin(2\theta-2\theta')(\sin3\theta-\sin3\theta') \bigg] \delta(\epsilon_{\bm{k}}-\epsilon_{\bm{k}'}),
\end{aligned}\\
\begin{aligned}
w_{s,\bm{k}\bm{k}'}^{(A)} = \ & \frac{\pi n V_i^3 k_F^2 \Delta}{4\epsilon_F^2} \bigg\{ s\sin(2\theta-2\theta') \\
& + \frac{v}{2\lambda k_F} \left[ \sin(\theta+2\theta') - \sin(2\theta+\theta') + \sin(2\theta-5\theta') + \sin(5\theta-2\theta') \right] \bigg\} \delta(\epsilon_{\bm{k}}-\epsilon_{\bm{k}'}).
\end{aligned}
\end{gather}

For the bilayer case, the scattering times become
\begin{gather}
\tau_1^{(0)} = \left( \frac{n_i V_i^2 (\epsilon_F^2+\Delta^2)}{4\lambda^2 k_F^2\epsilon_F} \right)^{-1}, \\
\tau_1^{(1)} = \tau_2^{(0)} = \tau_2^{(1)}
= \left( \frac{n_i V_i^2 (\epsilon_F^2+3\Delta^2)}{8\lambda^2 k_F^2\epsilon_F} \right)^{-1}, \\
\tilde{\tau} = \left(\Delta v \frac{n_i V_i^3}{32\lambda^3 k_F\epsilon_F}\right)^{-1}.
\end{gather}
We define $\tau$ as
\begin{gather}
\tau = \left( \frac{n_i V_i^2 (\epsilon_F^2+\Delta^2)}{4\lambda^2 k_F^2\epsilon_F} \right)^{-1},
\end{gather}
resulting in $\tau_1^{(0)} = \tau$ and $\tau_1^{(1)} = \tau_2^{(0)} = \tau_2^{(1)} = \gamma^{-1}\tau$, with the same $\gamma$ for the monolayer case.

The contributions to the second-order conductivity from valley $s$ are obtained from Eqs.~\eqref{eq:chi0} and \eqref{eq:chi1} at zero temperature as
\begin{gather}
\chi_{s,xxx}^{(0)} = -\chi_{s,xyy}^{(0)} = -\chi_{s,yxy}^{(0)} = -\chi_{s,yyx}^{(0)}
= s e^3\tau^2 \frac{2\lambda k_F^2(\epsilon_F^2+2\Delta^2)}{\pi\epsilon_F^3 \gamma} \operatorname{Re} \left[\frac{1}{1-i\omega\tau}\right], \\
\begin{aligned}
&\ \chi_{s,xxy}^{(1)} = \chi_{s,xyx}^{(1)} = \chi_{s,yxx}^{(1)} = -\chi_{s,yyy}^{(1)} \\
=&\ e^3\tau^2 (2\lambda k_F) \frac{\tau}{\tilde{\tau}} \frac{\lambda k_F^2 \Delta^2}{4\pi\epsilon_F^3 \gamma^2} \operatorname{Re} \left[ \frac{5}{(1-i\omega\tau)(1-i\omega\gamma^{-1}\tau)} - \frac{6}{1-i\omega\tau} \right].
\end{aligned}
\end{gather}
For the bilayer model, the ratio $\tau/\tilde{\tau}$ is given by
\begin{equation}
\frac{\tau}{\tilde{\tau}} = \Delta v \frac{k_F V_i}{8\lambda (\epsilon_F^2+\Delta^2)}.
\end{equation}
Similarly to the monolayer case, $\chi_{s,abc}^{(0)}$ from each valley cancels but $\chi_{s,abc}^{(1)}$ adds up, leading to the second-order conductivity
\begin{gather}
\label{eq:result-bilayer}
\chi_{xxy} = \chi_{xyx} = \chi_{yxx} = -\chi_{yyy} = -4e^3 v_F \frac{\tau^3}{\tilde{\tau}} \zeta(\omega) \equiv -\chi(\omega),
\end{gather}
where for the bilayer model the average velocity on the Fermi surface $v_F$ is
\begin{equation}
v_F = \frac{2\lambda^2 k_F^3}{\epsilon_F},
\end{equation}
and the dimensionless function $\zeta(\omega)$ becomes
\begin{gather}
\zeta(\omega) = \frac{\Delta^2}{8\pi\epsilon_F^2 \gamma^2} \operatorname{Re} \left[ \frac{6}{1-i\omega\tau} - \frac{5}{(1-i\omega\tau)(1-i\omega\gamma^{-1}\tau)} \right].
\end{gather}

\subsubsection{Linear conductivity and responsivity}

We also evaluate the linear conductivity $\sigma_{s,ab}^{(0)}$ for each valley $s$ with the spin degrees of freedom included [Eq.~\eqref{eq:linear_0}]:
\begin{equation}
\sigma_{s,ab}^{(0)} = e^2 \tau_{1\omega}^{(0)} \frac{\lambda^2 k_F^4}{\pi\epsilon_F} \delta_{ab}.
\end{equation}
Similarly to the monolayer case, we define
\begin{equation}
\sigma(\omega) = \sum_s \operatorname{Re} \sigma_{s,aa}^{(0)}
= e^2 \tau \frac{\lambda^2 k_F^4}{\pi\epsilon_F} \frac{1}{1+(\omega\tau)^2}.
\end{equation}
For low frequencies, the mobility is calculated from the relation $\sigma = ne\mu$ by
\begin{equation}
\mu = e\tau \frac{2\lambda^2 k_F^2}{\epsilon_F}.
\end{equation}
where the electron density is $n= k_F^2/\pi$.

For the bilayer case, the reduced current responsivity becomes
\begin{equation}
\eta_I = \frac{\chi(\omega)}{\sigma(\omega)}
= \frac{8\pi e\tau}{k_F} [ 1+(\omega\tau)^2 ] \zeta(\omega),
\end{equation}
and the reduced voltage responsivity is
\begin{equation}
\eta_V = \frac{\chi(\omega)}{\sigma(\omega)\sigma(0)}
= \frac{16\pi}{env_F} \left[ 1+(\omega\tau)^2 \right] \zeta(\omega).
\end{equation}

\section{Estimate of parameters}
\label{sec:values}

Parameters for a realistic evaluation are taken and estimated from experiments on monolayer and bilayer graphene.
For monolayer graphene on an hBN substrate, the imbalance of the potential energy within a sublattice is induced by the substrate and the lattice mismatch of graphene and hBN creates a superlattice with a periodicity of about 13\,nm.  In addition to the original (primary) Dirac points, the moir\'e superlattice produces new (secondary) superlattice Dirac points, located at a different energy \cite{Louie}.  Those two have different velocities and energy gaps: at ambient pressure, $v = 0.94\times 10^6\,\text{m/s}$ and $2\Delta = 30\,\text{meV}$ for the original Dirac point, and $v = 0.5\times 10^6\,\text{m/s}$ and $2\Delta = 20\,\text{meV}$ for the superlattice Dirac points \cite{LeRoy,Dean1}.
The size of the gap depends on the superlattice periodicity and it could be enhanced by interaction \cite{Song}.
We note that even at the original Dirac point, the velocity is slightly modified.
We consider the original Dirac points instead of the superlattice Dirac points.  The latter have the valley degeneracy of six and anisotropic Dirac cones whereas the valley degeneracy of the former is two.

The scattering time is estimated from a transport measurement with a high-quality sample \cite{Wang}.  At 300\,K, they observed the mobility $\mu\approx 90,000\,\text{cm}^{2}/(\text{V\,s})$ with the carrier density $n = 1\times 10^{12}\,\text{cm}^{-2}$, which amounts to the scattering time $\tau \approx 1.13\,\text{ps}$.  The corresponding frequency is $(2\pi\tau)^{-1} \approx 141\,\text{GHz}$ and the mean free path is $\ell \sim v_F\tau \approx 1.6\,\mu\text{m}$.
We note that the Fermi wavelength at $n=0.5\times 10^{12}\,\text{cm}^{-2}$ is $\lambda_F \approx 0.05\,\mu\text{m}$, which is still shorter than the mean free path: $\lambda_F \ll \ell$.
To estimate the impurity strength, we need to know the impurity concentration.  From transport and scanning tunneling microscopy/spectroscopy \cite{Dean2,Xue}, we approximate the impurity density as $n_i \approx 1\times 10^9\,\text{cm}^{-2}$, which results in $V_i \approx 2.77\times 10^{-13}\,\text{eV\,cm}^2$ ($nV_i \approx 277\,\text{meV}$ at $n=1\times 10^{12}\,\text{cm}^{-2}$).
Since the velocity is only slightly modified from the value without a superlattice structure, we employ the value $\lambda/v = a/4$ obtained from a tight-binding model, where $a$ is the carbon atom spacing, given by $a=1.42\,\text{\AA}$ \cite{graphene_review1,graphene_review2}.
Those values yields the ratio $\tau/\tilde{\tau} \approx 0.003$ at $n=1\times 10^{12}\,\text{cm}^{-2}$.
In the present calculations, we do not consider the temperature dependence of the parameters or the origin of scattering and inhomogeneity of samples for simplicity, although the scattering time becomes longer in lower temperatures.

For bilayer graphene, high-quality bilayer graphene grown by chemical vapor deposition (CVD) and detached on hBN realizes the mobility $\mu = 30,000\,\text{cm}^2/\text{(V\,s)}$ at 300\,K \cite{Schmitz}. An out-of-plane electric field (displacement field) $D$ opens a band gap of about $2\Delta = 100\,\text{meV}$ with $D\approx 1.1\,\text{V/nm}$ \cite{Lu}.  The coefficient $\lambda=(2m)^{-1}$ is determined by the effective mass $m\approx 0.033m_e$ ($m_e$: electron mass) and $v\approx 1\times 10^5\,\text{m/s}$ \cite{bilayer1,bilayer2}.  Evaluating at $n=1\times 10^{12}\,\text{cm}^{-2}$, we find the scattering time $\tau\approx 0.96\,\text{ps}$, corresponding to the frequency $(2\pi\tau)^{-1}\approx 166\,\text{GHz}$, and the mean free path $\ell \sim v_F\tau \approx 0.35\,\mu\text{m}$.  We assume the impurity concentration $n_i \approx 1\times 10^{9}\,\text{cm}^{-2}$, considering its high mobility, to obtain the impurity strength $V_i \approx 1.06\times 10^{-13}\,\text{eV\,cm}^2$ ($nV_i \approx 106\,\text{meV}$ at $n=1\times 10^{12}\,\text{cm}^{-2}$).  At the carrier density $n=1\times 10^{12}\,\text{cm}^{-2}$, we have the ratio $\tau/\tilde{\tau} \approx 0.018$.

\begin{figure}[b]
\centering
\includegraphics[width=0.85\hsize]{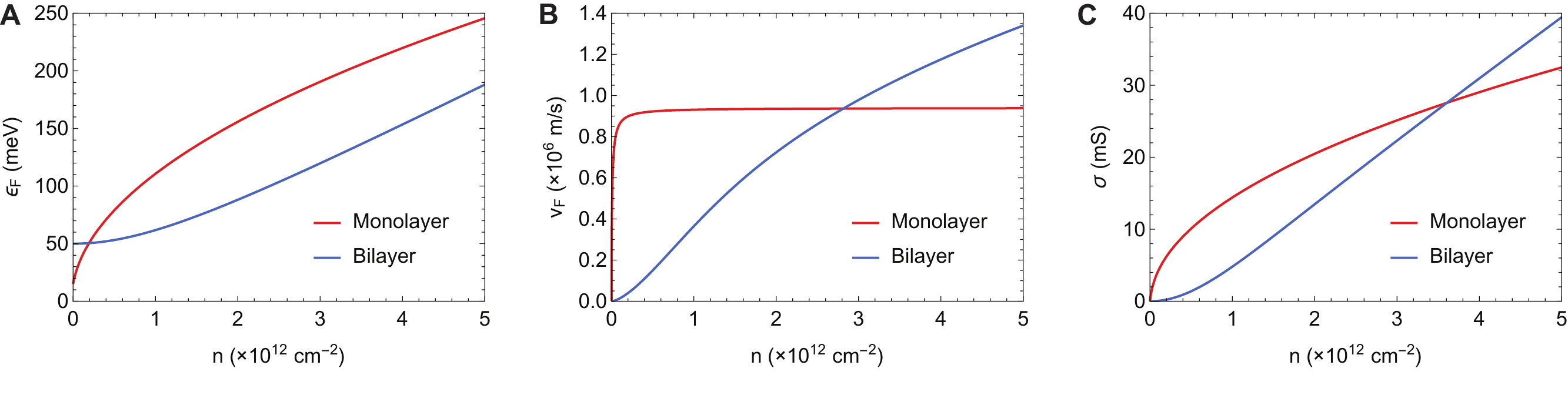}
\caption{
\textbf{Carrier density dependence of the material properties.}
(\textbf{A}) Fermi energy $\epsilon_F$, (\textbf{B}) Fermi velocity $v_F$, and (\textbf{C}) linear DC conductivity $\sigma$ at zero temperature. Refer to Sec.~\ref{sec:values} for the values of the parameters.
}
\label{fig:result_density}
\end{figure}

\begin{figure}
\centering
\includegraphics[width=0.67\hsize]{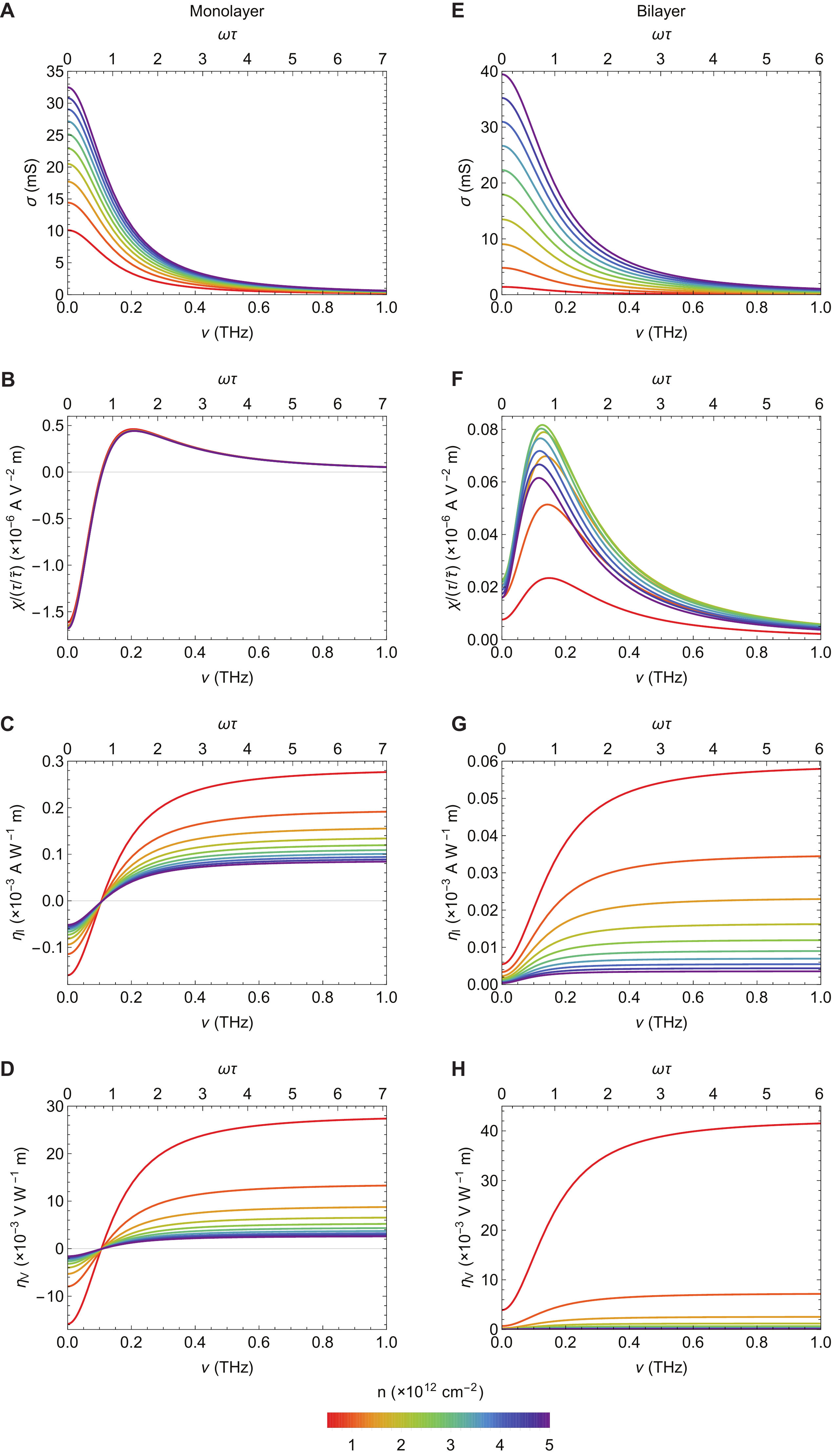}
\caption{
\textbf{Frequency dependence of the response.}
(\textbf{A} and \textbf{D}) Linear conductivity $\sigma$, (\textbf{B} and \textbf{F}) second-order conductivity $\chi$, (\textbf{C} and \textbf{G}) reduced current responsivity $\eta_I$, and (\textbf{D} and \textbf{H}) reduced voltage responsivity $\eta_V$.  Panels (A) to (D) correspond to the monolayer model and (E) to (H) to the bilayer model.  The carrier density is changed from $0.5\times 10^{12}\,\text{cm}^{-2}$ (red) to $5\times 10^{12}\,\text{cm}^{-2}$ (purple), and the temperature is set to be zero.  Refer to Sec.~\ref{sec:values} for the values of the parameters.
}
\label{fig:result_freq}
\end{figure}

\begin{figure}
\centering
\includegraphics[width=0.7\hsize]{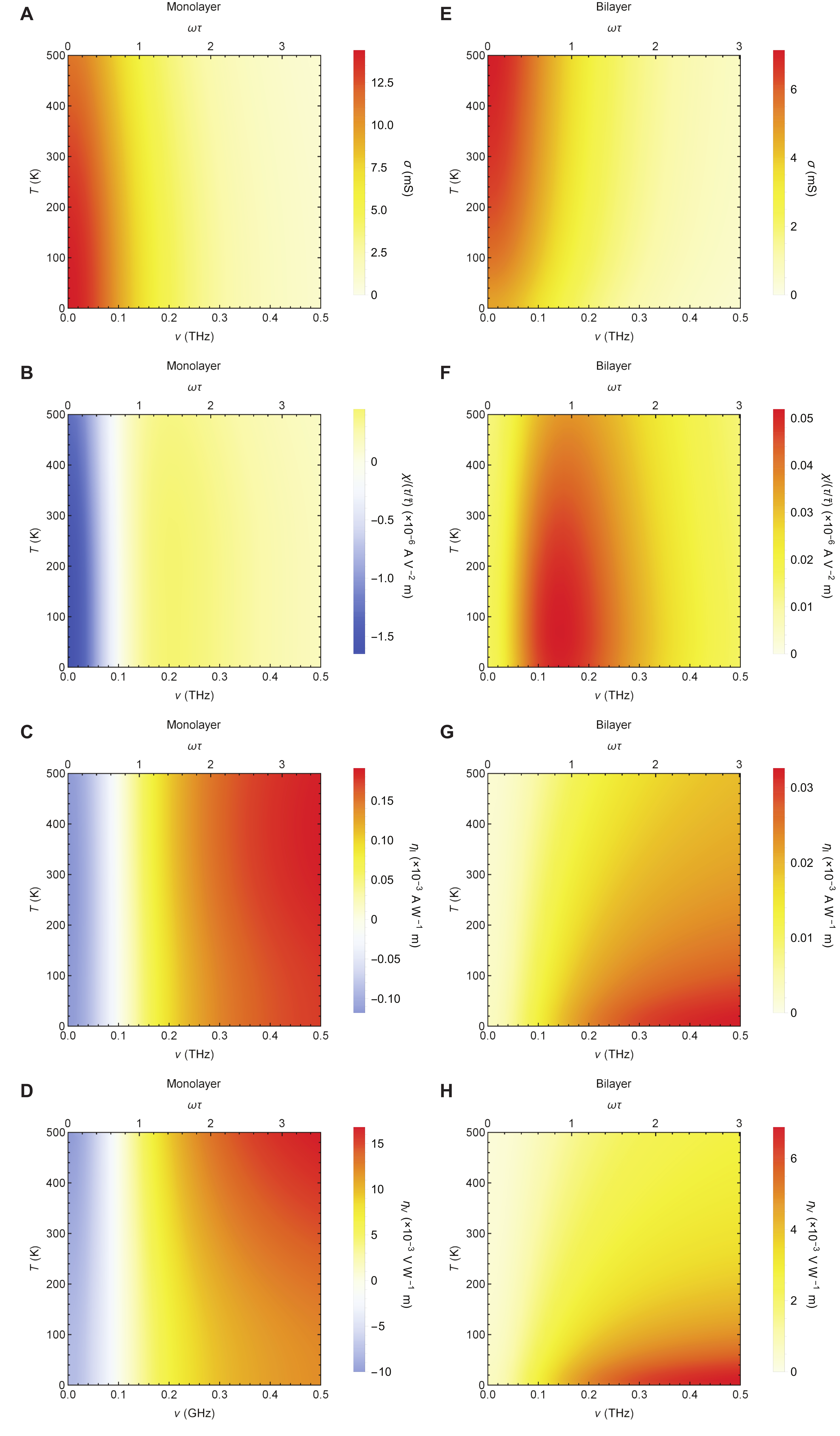}
\caption{
\textbf{Frequency and temperature dependence of the response.}
(\textbf{A} and \textbf{E}) Linear conductivity $\sigma$, (\textbf{B} and \textbf{F}) second-order conductivity $\chi$, (\textbf{C} and \textbf{G}) reduced current responsivity $\eta_I$, and (\textbf{D} and \textbf{H}) reduced voltage responsivity $\eta_V$.  
The carrier density is fixed at $n=1\times 10^{12}\,\text{cm}^{-2}$.
Panels (A) to (D) correspond to the monolayer model and (E) to (H) to the bilayer model.
We assume that the scattering time $\tau$ is constant independent of temperature; see Sec.~\ref{sec:values} for the values of parameters.
}
\label{fig:result_map}
\end{figure}

\begin{figure}
\centering
\includegraphics[width=0.7\hsize]{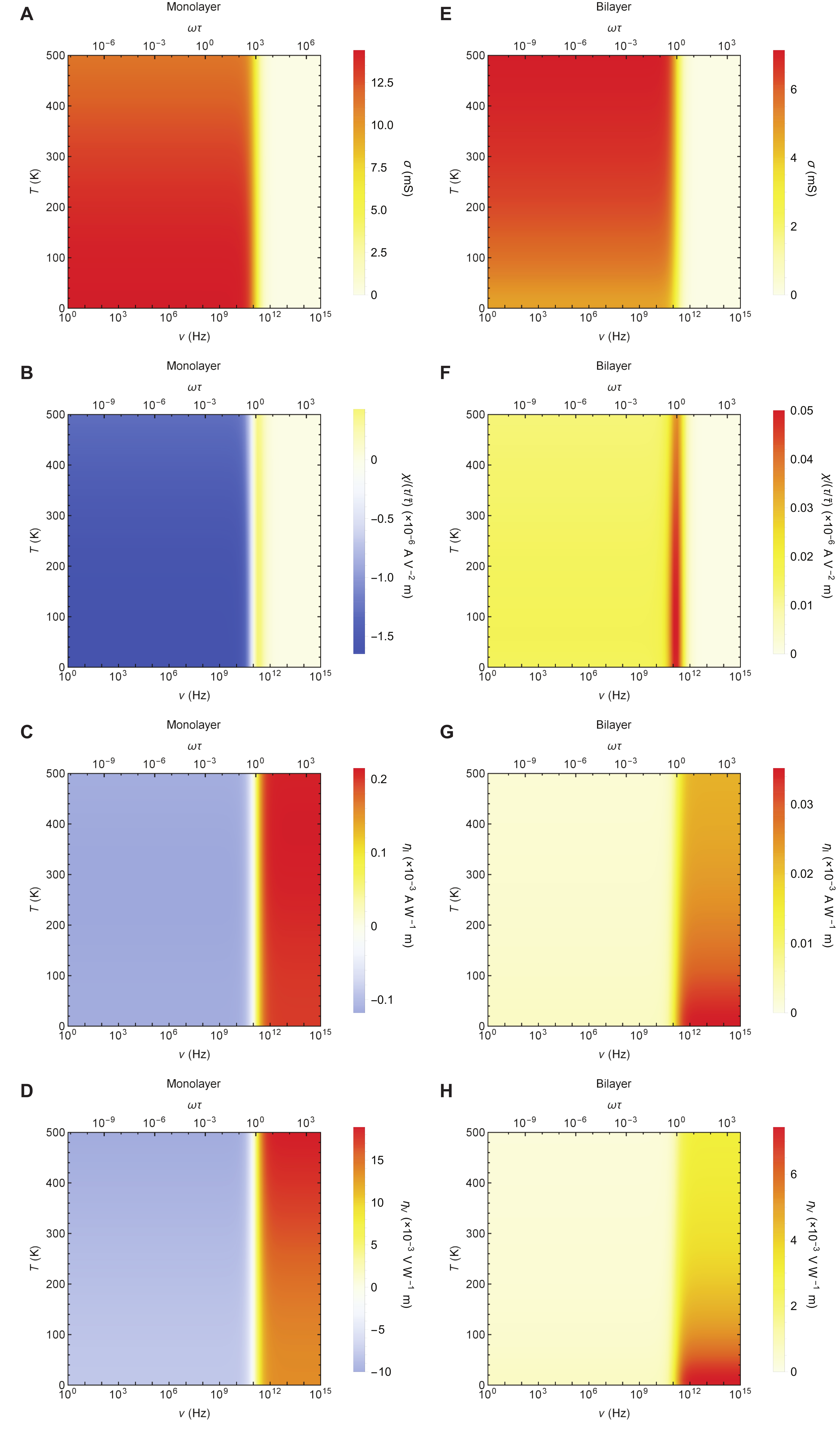}
\caption{
\textbf{Frequency and temperature dependence of the response with a logarithmic scale for the frequency axis.}
(\textbf{A} and \textbf{E}) Linear conductivity $\sigma$, (\textbf{B} and \textbf{F}) second-order conductivity $\chi$, (\textbf{C} and \textbf{G}) reduced current responsivity $\eta_I$, and (\textbf{D} and \textbf{H}) reduced voltage responsivity $\eta_V$.  
The same results as Fig.~\ref{fig:result_map} are plotted with the frequency in a logarithmic scale. 
}
\label{fig:result_map_log}
\end{figure}

The carrier density dependence of $\epsilon_F$, $v_F$, and $\sigma(0)$ is shown in Fig.~\ref{fig:result_density}, and the frequency dependence of $\sigma$, $\chi$, $\eta_I$, and $\eta_I$ is depicted in Fig.~\ref{fig:result_freq}.
The frequency and temperature dependence of the response are given in Figs.~\ref{fig:result_map} and \ref{fig:result_map_log}.
We note that the scattering times used for the calculations are those at 300\,K.  The scattering time becomes longer in lower temperature, which makes the linear conductivity $\sigma$, the second-order conductivity $\chi$, and the reduced current responsivity $\eta_I$ larger, whereas the low- and high-frequency values of the reduced voltage responsivity $\eta_V$ barely change.

\clearpage

\clearpage

\clearpage

\clearpage

\section{Surface state of a topological insulator}

A surface state of a topological insulator also lacks inversion and thus hosts the rectification effect \cite{Ganichev}.  Here we take an example of the surface state of the topological insulator Bi$_2$Te$_3$ \cite{TIsurface,Ganichev}, where the Hamiltonian is
\begin{equation}
H(\bm{k}) = v(k_x\sigma_y - k_y\sigma_x) + \frac{\lambda}{2}(k_+^3+k_-^3)\sigma_z,
\end{equation}
leading to the energy spectrum
\begin{equation}
\epsilon_{\bm{k}} = \sqrt{v^2k^2 + \lambda^2 k^6 \cos^2 3\theta}.
\end{equation}
The origin $\bm{k}=0$ is located at $\Gamma$ point and the $k_x$ axis is parallel to $\Gamma K$ line. 
Unlike the previous two cases based on graphene, it has a single Dirac spectrum centered at $\Gamma$ point.  Furthermore, the topology of the wavefunction in the bulk prohibits a mass gap of the surface states.  In this case, a surface breaks inversion by itself and the symmetry is $C_{3v}$.

For the present Hamiltonian, short-ranged impurities without momentum dependence do not produce finite $w^{(A)}$, nor does the second-order response.  This issue can be circumvented instead by considering Coulomb impurities, where the impurity potential is inversely proportional to the distance from the impurity position: $V(\bm{r})\propto \sum_j |\bm{r}-\bm{r}_j|^{-1}$.
Assuming that the hexagonal warping of the Fermi surface is small, we have the matrix element of the Coulomb impurity potential on the Fermi surface as
\begin{equation}
V_{\bm{k}\bm{k}'} = \frac{V_c}{\left|\sin\left(\dfrac{\theta-\theta'}{2}\right)\right|}.
\end{equation}
Then, we obtain the scattering rates
\begin{gather}
w_{\bm{k}\bm{k}'}^{(S)} = \frac{2\pi n_i V_c^2}{\tan^2\left(\dfrac{\theta-\theta'}{2}\right)} \delta(\epsilon_{\bm{k}}-\epsilon_{\bm{k}'}), \\
w_{\bm{k}\bm{k}'}^{(A)} \approx \frac{8\pi n_i V_c^3 \epsilon_F^3}{v^5} \lambda c_1 \operatorname{sgn}\left[ \sin\left(\frac{\theta-\theta'}{2}\right) \right] \cos\left(\frac{\theta-\theta'}{2}\right) (\cos3\theta-\cos3\theta') \delta(\epsilon_{\bm{k}}-\epsilon_{\bm{k}'}),
\end{gather}
with the numerical factor $c_1 \approx 0.93$ obtained from the numerical integration over momenta.

Since there is a single Fermi surface and time-reversal symmetry relates the Fermi surface with itself, $\chi^{(0)}$ vanishes with the single Fermi surface.  A finite second-order conductivity can be found in $\chi^{(1)}$ as follows:
\begin{align}
\chi_{xxy}^{(1)} = \chi_{xyx}^{(1)} = \chi_{yxx}^{(1)} = -\chi_{yyy}^{(1)}
\approx e^3 v \frac{\tau^3}{\tilde{\tau}} \operatorname{Re} \left[ \frac{c_1}{1-i\omega\tau} + \frac{c_2}{(1-i\omega\tau)^2} \right].
\end{align}
Here, we assume for simplicity the constant scattering time $\tau\approx (2\pi D_0 n_i V_c^2)^{-1}$ obtained from $w^{(S)}$, and the scattering time originated from $w^{(A)}$ is defined by
\begin{equation}
\tilde{\tau}^{-1} = \lambda \frac{2n_iV_c^3 \epsilon_F^4}{\pi v^7}.
\end{equation}
The two numerical factors $c_1 \approx 0.87$ and $c_2 \approx 0.14$ are evaluated by numerical integrations.

\end{document}